\documentclass[useAMS,usenatbib]{mn2e}
\usepackage{epsf}
\usepackage{amssymb}
\usepackage[usenames]{color}

\newcommand {\bc}{\begin {center}}
\newcommand {\ec}{\end {center}}
\newcommand {\be}{\begin {equation}}
\newcommand {\ee}{\end {equation}}
\newcommand {\beq}{\begin {eqnarray}}
\newcommand {\eeq}{\end {eqnarray}}

\def\plotone#1{\centering \leavevmode
\epsfxsize=\columnwidth \epsfbox{#1}}

\def\plottwo#1#2{\centering \leavevmode
\epsfxsize=1.\columnwidth \epsfbox{#1}\hfil 
\epsfxsize=1.\columnwidth \epsfbox{#2}}

\def\plotwide#1{\centering \leavevmode
\epsfxsize=1.99\columnwidth \epsfbox{#1}}

\def\disp {\displaystyle}

\title[Constraints on the ICM velocity power spectrum from the X-ray
  lines]{Constraints on the ICM velocity power spectrum from the X-ray lines width and shift}
\author[Zhuravleva et al.]{I.Zhuravleva$^{1}$\thanks{izhur@mpa-garching.mpg.de},
  E.Churazov$^{1,2}$, A.Kravtsov$^{3,4}$, R.Sunyaev$^{1,2}$\\ \\
$^{1}$MPI f\"ur Astrophysik, Karl-Schwarzschild str. 1, Garching, 85741, Germany\\
$^{2}$Space Research Institute, Profsoyuznaya str. 84/32, Moscow, 117997, Russia\\
$^3$Department of Astronomy and Astrophysics, University of Chicago,
  5640 South Ellis Avenue, Chicago, IL 60637, USA\\
$^4$Kavli Institute for Cosmological Physics and Enrico Fermi
  Institute, University of Chicago, Chicago, IL 60637, USA
  }

\begin{document}

\date{Accepted .... Received ...}

\pagerange{\pageref{firstpage}--\pageref{lastpage}} \pubyear{2009}

\maketitle

\label{firstpage}

\begin{abstract}
Future X-ray observations of galaxy clusters by high spectral
resolution missions will provide spatially resolved measurements of
the energy and width for the brightest emission lines in the
  intracluster medium (ICM) spectrum. In this paper
  we discuss various ways of using these high resolution data to
  constrain velocity power spectrum in galaxy clusters. We argue that
variations of these quantities with the projected distance $R$ in
cool core clusters contain important information on the velocity
field length scales (i.e. the size of energy-containing eddies) in the ICM. The effective
length $l_{\rm eff}$ along the line of sight, which provides dominant
contribution to the line flux, increases with $R$, allowing one
to probe the amplitude of the velocity variations at different spatial
scales. In particular, we show that the width of the line as a
function of $R$ is closely linked to the structure function of the 3D
velocity field. Yet another easily obtainable proxy of the velocity
field length scales is the ratio of the amplitude of the projected
velocity field (line energy) variations to the dispersion of the
velocity along the line of sight (line width). Finally the projected velocity field can be easily converted into 3D velocity field, especially for clusters like Coma with an extended flat core in the surface brightness. Under assumption of a
homogeneous isotropic Gaussian 3D velocity field we derived simple expressions
relating the power spectrum of the 3D velocity field (or structure
function) and the observables. We illustrate the sensitivity of these
proxies to changes in the characteristics of the power spectrum for a
simple isothermal $\beta$-model of a cluster. The uncertainties in the
observables, caused by stochastic nature of the velocity field, are
estimated by making multiple realizations of the random Gaussian
velocity field and evaluating the scatter in observables. If large scale motions are present in the ICM these uncertainties may dominate the statistical errors of line width and shift measurements.    
\end{abstract}
\section{Introduction}
\begin{figure}
\plotone{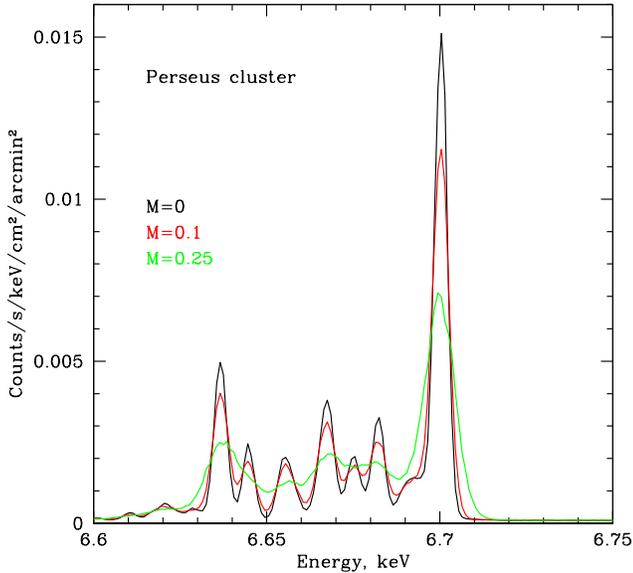}
\caption{ Simulated spectrum of the Perseus cluster core (central 130 kpc), calculated assuming different Mach numbers of the
  turbulence. At each point along the line of sight the locally
  emitted spectrum is
  calculated using Astrophysical Plasma Emission Code
  (APEC, \citet{Smi01}) model using known radial density and
  temperature dependencies. The lines are then broadened by the thermal
  and turbulent motions using eq. (\ref{eq:deltaE}) (see
  \S\ref{sec:dislines}) and the whole spectrum is projected along the
  line of sight.
  The case of pure thermal broadening of lines is shown in
  black. This plot shows that (i) turbulent broadening is important
  even for the modest level of turbulence and (ii) lines overlap with
  each other, especially if turbulent motions are present.
\label{fig:perlines}
}
\end{figure}

Properties of gas motions in the hot intracluster medium (ICM) are still
little known. It is believed that turbulent motions are driven when matter
accretes onto the filaments or during shocks in the hot gas.
Turbulence transfer kinetic energy injected on large scales $L \sim$ Mpc to small (unknown)
dissipative scales $l$. These two scales are connected with a cascade of
kinetic energy, which occurs over inertial range \citep{Kol41,Lan66}.

Knowing the properties of gas motions in clusters, we would be able to
address a number of important question, e.g., what is the bias in the
cluster mass
measurements based on hydrostatic equilibrium and whether the bias is
  due to the 
motions alone or due to the clumping in gas density \citep[see, e.g.,
][]{Ras06,Nag07,Jel08,Lau09}, what is the ICM turbulent heating
rate in clusters \citep[e.g.][]{Chu08} and what is the role of gas
motions in particle acceleration \citep[see, e.g. ][]{Bru06,Bru11}.

Properties of turbulence in galaxy clusters were studied by means
of numerical simulations \citep[e.g.,][]{Dol05,Cas05,Nor99,Iap11,Vaz11}.
However despite the good ``global" agreement between all simulations,
the results on turbulent motions are still controversial, 
mainly due to insufficient resolution of simulations and, in particular, low Reynolds number (effective $Re<1000$ in cosmological
simulations) and other
numerical issues \citep[see, e.g.,][]{Kit09,Dob03,Ber09}.

Current generation of X-ray observatories cannot provide robust direct
measurements of turbulence in the ICM. Only XMM RGS grating can provide weak upper
limits on velocity amplitude in cool core clusters \citep{San11}. Indirect
indications of the ICM turbulence come from measurements of the resonant
scattering effect \citep[e.g.][]{Chu04,Wer09}, from measurements of
pressure fluctuations \citep{Sch04} or surface brightness fluctuations \citep{Chu12}

Future X-ray observatories, such as {\it Astro-H}\footnote{http://astro-h.isas.jaxa.jp/} and {\it
ATHENA}\footnote{http://sci.esa.int/ixo}, with their high energy resolution will allow us to measure
shifts and broadening of individual lines in spectra of galaxy clusters
with high accuracy. Combination of direct measurements of velocity
amplitudes with indirect measurements via resonant scattering will
give us constraints on anisotropy of motions
\citep{Zhu11}. X-ray polarimetric measurements can also
provide information on gas motions perpendicular to the
line of sight \citep{Zhu10}.

Here we discuss the possibility of getting information about
 the length scales of
gas motions (e.g. the size of energy-containing eddies). We discuss various ways to constrain structure function and power
spectrum of gas motions via measurement of the projected velocity
(shift of the line centroid) and
the velocity dispersion (broadening of the line) as a function of projected distance from the
cluster center. These ideas are illustrated with a very
  simple model of a galaxy cluster. An application of our methods to 
  simulated galaxy clusters will be considered in future work. 

\begin{figure*}
\plotwide{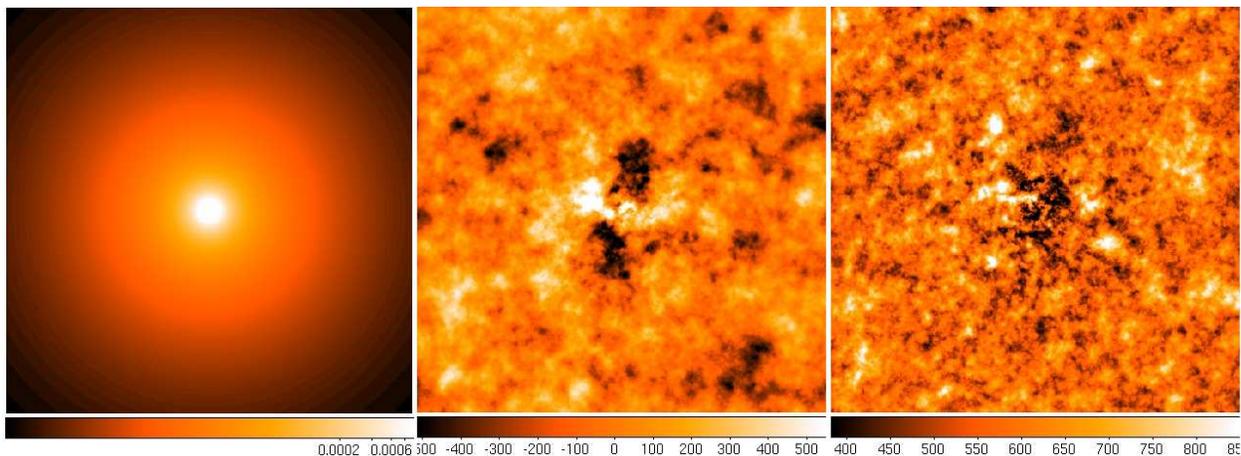}
\caption{Maps of observables. Left: surface brightness in line (arbitrary units);
  middle : emissivity-weighted projected velocity field in km/s (centroid shift of the
  line); right: velocity
  dispersion in km/s (line broadening). Here we consider only the velocity component
  along the line of sight. We assume $\beta$-model of the emissivity
  with $\beta=0.6$ and $r_c=10$ kpc and a random, uniform and homogeneous Gaussian velocity field with cored power 3D
  power spectrum (eq. \ref{eq:ps3d}) with $k_m=0.005$ kpc$^{-1}$ and Kolmogorov
  slope $\alpha=-11/3$  (see Section 2). The size of
  each map is 1Mpc$\times$1Mpc.
\label{fig:maps}
}
\end{figure*}
 
A similar problem of obtaining the structure function of turbulence
from spectral observations has been addressed in the studies of
Galactic interstellar medium. In particular, it was shown that the
width of molecular spectral lines increases with the size of a cloud
\citep[see e.g.][]{Mye78,Hey04,Hey09}. This correlation was
interpreted in terms of turbulent velocity spectrum \citep{Lar81}.  A
way to constrain structure function of turbulence in the Interstellar
Medium (ISM) by means
of the velocity centroids (projected mean velocity) measurements was first considered by
\citet{Hoe51} and \citet{Mun58} \citep[see also
    ][]{Kle83,Kle85}.  Currently several different flavors of the
  velocity centroids method are used for studies of the ISM turbulence \citep[see
    e.g.][]{Esq07}. More advanced techniques, such as Velocity Channel
  Analysis (VCA) and Velocity Coordinate Spectrum (VCS) \citep[see
    e.g.][]{Laz00,Laz08,Che09}, were developed and applied to the ISM data
  in the Milky
  Way and other galaxies \citep[see e.g.][]{Pad09,Sta01,Che10}.  Few
  other methods were also used, among them are the Spectral
  Correlation Function (SCF) \citep{Ros99,Pad03} and the
  Principal Component Analysis (PCA) \citep{Bru03}.

The ISM turbulence is often supersonic and compressible
\citep[e.g.][]{2004ARA&A..42..211E}. This leads to (i) large shifts
in the energy centroid of the line compared to the thermal broadening and
(ii) large amplitude of the gas density fluctuations. At the same time
individual lines, which are used to study the ISM turbulence (e.g. 21
cm line of HI or CO lines) are often well separated from other
emission lines. The regions under study often have very irregular
structure on a variety of spatial scales. The analysis therefore is
usually concentrated on separation of the velocity and density
fluctuations in the observed data, while the thermal broadening can
often be neglected.

In contrast, in galaxy clusters the gas motions are mostly
subsonic. The detection of the gas motions is still possible, since we
deal with the emission lines of ions of heavy elements like e.g. Fe,
Ca or S. The atomic weights of these elements are large (e.g. for Fe it
is 56) and this drives pure thermal broadening of lines down (see
Fig. \ref{fig:perlines} and Section \ref{sec:dislines}). The
brightest lines in spectra of galaxy clusters are often very close
to each other. For example, in the vicinity of the He-like iron line
at 6.7 keV there are forbidden and intercombination lines and a number
of satellite lines, energy separation between which is of the order
few tens eV (Fig. \ref{fig:perlines}). Density of clusters often have a regular
radial structure with relatively small amplitude of stochastic density
fluctuations. Analysis of X-ray surface brightness fluctuations in
Coma cluster shows that density fluctuations are $\sim 2-10$ per
cent \citep{Chu12}. Also hydrodynamical simulations of cluster
formation predict very small clumping factors within $r_{500}$
\citep[see e.g.][]{Mat99,Nag11}. Therefore, to the first approximation, the
contribution of density fluctuation in galaxy clusters can be
neglected (see Section \ref{sec:denfl} for details), while global
radial dependence has to be taken into account (Section \ref{sec:sf}). Another
characteristic feature of X-ray observations is the importance of the
Poisson noise, related to the counting statistics of X-ray
photons. If one deals with the clusters outskirts the high energy
resolution spectra will be dominated by the Poisson noise even for
large area future telescopes, like e.g. ATHENA\footnote{http://sci.esa.int/ixo}.
Finally, one
can mention, that the
effects of self-absorptions can potentially be important in clusters. Galaxy
clusters are transparent in X-rays in continuum and in most of the
lines. However, some strong lines can have an optical depth $\sim$
few units. Therefore if one measures the width of the optically thick
line distortions due to resonant scattering effect should be taken
into account \citep[see e.g.][]{Chu10,Wer09,Pla12,Zhu11}.

Presence of several closely spaced emission lines, modest level of
turbulence (i.e. modest ratio of the turbulent and thermal
broadenings), lack of very strong stochastic density fluctuations on
top of a regular radial structure, and often strong level of Poisson
noise affect the choice of the simplest viable approach to relate
future observables and most basic characteristics of the ICM gas
velocity field. 

 The fact that lines in spectra of galaxy clusters are very close
to each other (e.g. Fig. \ref{fig:perlines}) and the line ratio is temperature
dependent can be circumvented by estimating the mean shift
and broadening with direct fitting of the projected
spectra with the plasma emission model (possible multi-temperature 
model). While the small thermal broadening of lines from heavy
elements helps to extend the applicability of VCA/VCS techniques
into subsonic regime \citep[see][]{Esq03,Laz03,Che10}, limited spectral resolution of the next
generation of X-ray bolometers (e.g. ${\rm FWHM}\sim 5$eV for
ASTRO-H) reduces the measured inertial range in the velocity
domain. Direct application of SCF and PCA methods to galaxy
clusters can also be challenging, especially when the spectra are
dominated by the Poisson noise. These problems should be
alleviated with missions like ATHENA, having very large effective
area and an energy resolution of $\sim 2.5$ eV.

Below we suggest to use the simplest ``centroids and broadening''
approach as a first step in studying the ICM turbulence. This approach
assumes that at any given position one fits the observed spectrum with
a model of an optically thin plasma (including all emission lines) and
determines the velocity centroid and the line broadening. This
reduces the whole complexity of X-ray spectra down to two numbers -
shift and broadening. Simple analysis of existing hydrodynamic
simulations of galaxy clusters shows that this approximation does a
good (although not perfect) job in describing the profiles of emission
lines \citep[see e.g. Fig.2 in][]{Ino03}. At the same time this
approach is the most effective in reducing the Poison noise in the
raw measured spectra. We show below that in spite of its simplicity
this approach provided an easy way to characterize the most basic
properties of the ICM velocity field.

Clearly, more sophisticated methods developed for the
ISM turbulence \cite[e.g.][]{Laz09} will eventually be adopted to
the specific characteristics of the ICM turbulence and X-ray spectra,
potentially providing a more comprehensive description of the ICM
turbulence, once the data of sufficient quality become available.

 The structure of the paper is as follows. In Section 2 we
  describe and justify models and assumptions we use in our
  analysis. In Section 3 we specify observables which can be
  potentially measured and their relation to the 3D velocity power
  spectrum. Section 4 shows the relation between observed velocity
  dispersion and the structure function of the velocity field. A way to
  constrain length scales (size of the energy containing eddies) of
  motions using observed projected velocity is presented in Section
  5. Method to recover 3D velocity PS from 2D projected velocity field is
  discussed in Section 6. Discussions and conclusions are in Sections 7
  and 8 respectively.
\begin{figure}
\plotone{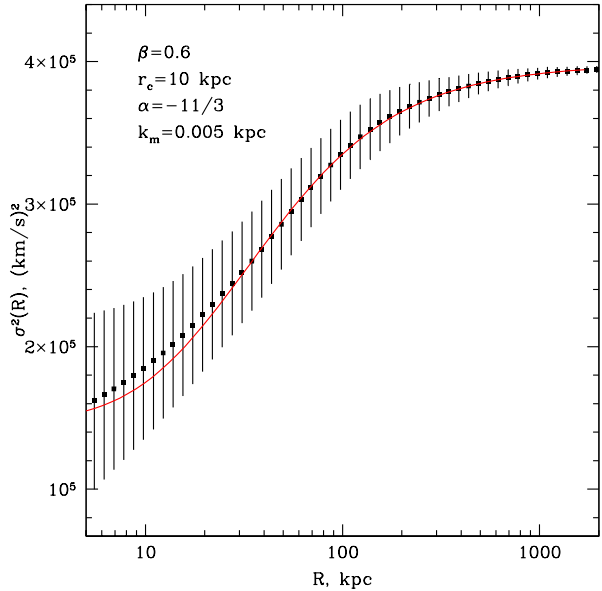}
\caption{Expected dependence of the line-of-sight velocity
  dispersion (line width) on the projected
  distance $R$ from the cluster center. The increase of the effective
  length along the line of sight with the projected distance $R$
  implies that larger and larger eddies contribute to the observed
  line broadening.
We assume a $\beta$ model
  of the density distribution in an isothermal galaxy cluster
  (eq. \ref{eq:beta}) and a cored 3D power law model of the velocity power
  spectrum with slope $\alpha$ and break at $k=k_m$ (eq. \ref{eq:psmodel}). Parameters are
  shown on the top left corner of the plot. Black dots with error bars: velocity 
  dispersion with uncertainties in a single realization, obtained by averaging over 100 realizations (the left part of
  eq. \ref{eq:SPS}), red curve: analytic expression of the velocity
  dispersion (the right hand side of eq. \ref{eq:SPS}, see Appendix B). 
\label{fig:sfth}
}
\end{figure}

\begin{figure}
\plotone{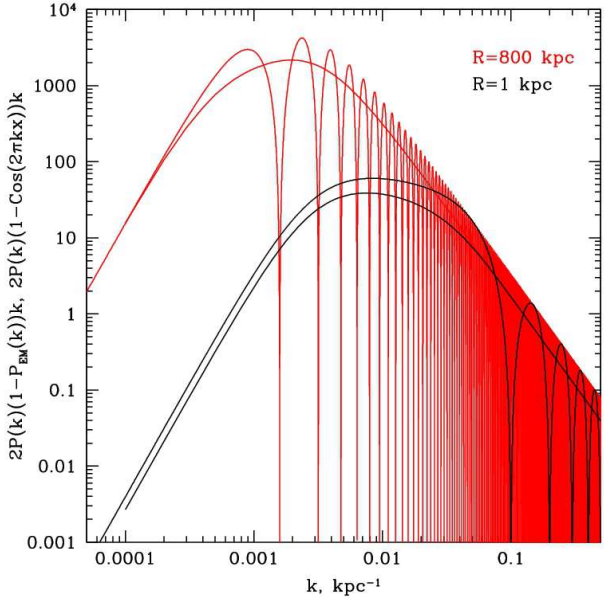}
\caption{Integrand from eq. \ref{eq:SF1d} multiplied by $k$ (oscillating function) and integrand from
  eq. \ref{eq:S1d} multiplied by 2k (non-oscillating function)
  calculated for projected distance $R=1$ kpc (black curves)  and  $R=800$ kpc from the
  center (red curves). The scales $\left(l\propto \disp\frac{1}{k}\right)$ where these functions have maxima
  provide dominant contribution to the line-of-sight velocity
  dispersion and structure function. The shift of the dominant scale
  to smaller $k$ at large projected distances shows that different
  parts of the power spectra are probed with $\sigma^2(R)$ measured at
  different $R$. The shift of the dominant scale in the $SF(l_{\rm
    eff}(R))$ traces the similar shift for $\sigma^2(R)$. This
  suggests that $\sigma^2(R)$ measurements can be used as a proxy for
  the structure function at a range of scales.
\label{fig:integr}
}
\end{figure}

\section{Basic assumptions and models}
 We consider a spherically symmetric galaxy cluster, which has a peaked X-ray emissivity
  profile. The electron number density is
described by the $\beta$-model profile
\be
n_e(r)=\frac{n_0}{\left[1+\disp\left(\frac{r}{r_c}\right)^2\right]^{\frac{3}{2}
\beta}},
\label{eq:beta}
\ee
where $n_0$ is the electron number density in the cluster center (normalization) and $r_c$ is the core
radius. The $\beta$-model provides a reasonably good description of observed
surface brightness \citep{Cav78} in the central regions of galaxy clusters. At large
radii $\beta-$ model is not a good description of surface brightness \citep[see e.g.][]{Vik06}. However the
simplicity of the model allows us to illustrate the method and make
analytical calculations.

We have chosen $\beta=0.6$ and $r_c=10$ kpc
for demonstration of our analysis. Parameter $\beta$ usually varies $0.4<\beta<1$ for galaxy clusters
\citep[see, e.g.,][]{Che07}, herewith a large fraction of clusters
have $\beta\sim 0.6$. $r_c$ can vary from few kpc to few hundreds
kpc. In order to better illustrate the main idea of the method,
  we considered cool-core clusters with small core radius $r_c$ (it is
  necessary to have a gradient of surface brightness down to the
  smallest possible projected distances, see Section 4 for details).

We describe the line-of-sight component of the 3D velocity field
as a Gaussian isotropic and homogeneous random field. This allows
us to gauge whether useful statistics could in principle be
obtained. However, there is no guarantee that this assumption
applies to the velocity field in real
clusters. E.g. \citet{Esq07} using numerical simulations have
shown that in the case of supersonic turbulence in the ISM, the
non-Gaussianity causes some of the statistical approaches (based
on the assumption of Gaussianity) to fail. The same authors
demonstrated that for the subsonic turbulence the Gaussianity
assumption holds much better. This is encouraging since in
clusters we expect mostly subsonic turbulence. Nevertheless the
methods discussed here require numerical testing using galaxy
clusters from cosmological simulations. We defer these tests for
future work.

Power spectrum (PS) of velocity field is described by a cored power law
model \footnote {Here and below we adopt the relation between a
  wavenumber $k$ and a spatial scale $x$ as 
  $k=1/x$ (without a factor $2\pi$).}
\be
P_{\rm 3D}(k_{\rm x},k_{\rm y},k_{\rm z})=\frac{B}{\left(1+\disp\frac{k_x^2+k_y^2+k_z^2}{k_{\rm
      m}^2}\right)^{\alpha/2}},
\label{eq:psmodel}
\ee
where $k_{\rm m}$ is a break wavenumber (in our simple model
$k_{\rm m}$
characterizes the injection scale),
$\alpha$ is a slope of the PS at $k>k_{\rm m}$ (inertial range) and
$B$ is the PS normalization, which is defined so
that the characteristic amplitude $A$ of velocity fluctuations at $k=k_{\rm ref}$ is fixed, i.e.
\be
B=\frac{A^2}{4\pi k_{\rm ref}^3 P_{\rm 3D}(k_{\rm ref})}.
\ee 
The cored power law model of the PS is a convenient description of the
PS for
analytical calculations and at the same time resembles widely used broken power law model.

 Now let us specify the choice of parameters $\alpha$ and $k_m$ in the
model of the velocity PS. Cluster mergers, motions of galaxies and AGN
feedback lead to turbulent motions with eddy sizes ranging from Mpc
near the virial radius down to few tens of kpc near the
cluster core \citep[see, e.g.,][]{Sun03}. For our analysis we vary
injection scales from 20 kpc ($k_m=0.05$ kpc$^{-1}$) to 2000 kpc
($k_m=0.0005$ kpc$^{-1}$).

Parameter $\alpha$ - the slope of the PS - can be selected using standard arguments.
 If most of the kinetic energy is on large scales
(injection scales), i.e. the characteristic velocity $V$ decreases
with $k$ then the power spectrum $PS=V^2/k^3\propto k^{\alpha}$ with
$\alpha\le -3$. At the same time the turnover time of large eddies
should not be larger than the turnover time of small eddies, i.e
$t=l/V=1/(kV)$ is a decreasing function of the wavenumber $k$. Therefore,
$V\propto k^{\gamma}$ with $\gamma\ge -1$ and $PS\propto V^2/k^3\propto
k^{\alpha}$ with $\alpha\ge -5$. So we expect the slope of the PS to
be in the range $-5\le\alpha\le-3$. We will use $\alpha=-11/3$
(slope of the Kolmogorov PS), $\alpha=-4$ and $\alpha=-4.5$.

\section{Observables and 3D velocity power spectrum}

Gas motions in relaxed galaxy clusters are predominantly subsonic, and to the first approximation the width and
centroid shift of lines measured with X-ray observatories
 contain most essential information on the ICM velocity field
  \citep[see, e.g.,][]{Ino03,Sun03}. That is, we
 have information about:

(i) surface brightness in lines (Fig. \ref{fig:maps}, left
panel), i.e. 
$I(x,y)\propto\int n_e^2(x,y,z) dz$ if one assumes isothermal galaxy
cluster (effects of non-constant temperature and abundance of
elements are discussed in Section 6),

(ii) emissivity-weighted projected velocity (Fig. \ref{fig:maps}, middle
panel) $ V_{\rm 2D}(x,y) =\disp \frac {\int V(x,y,z) n_e^2(z)
  dz}{\int n_e^2(z) dz}$ (in practice measured projected velocity is
averaged over some finite solid angle, see Section 5),

(iii) emissivity-weighted velocity dispersion (Fig. \ref{fig:maps}, right
panel) $\sigma(x,y)=\sqrt{\langle V^2(x,y,z)\rangle_z
-\langle V(x,y,z)\rangle^2_z} = \sqrt{\langle V^2(x,y,z)\rangle_z
- V^2_{\rm 2D}(x,y)}$,

\noindent where $n_e$ denotes the number electron density and $V$ is a
velocity component along the line of sight. Here $\langle \rangle_z$
denotes emissivity-weighted averaging along the
line of sight, which we assume to be along $z$ direction.

Relation between the PS of the 3D velocity field, the 2D projected velocity
and the velocity dispersion for a line of sight with projected coordinates $(x,y)$ are the following:

\be
\langle |\hat V_{\rm 2D}(k_x, k_y,x,y)|^2\rangle=\int P_{\rm 3D}(k_x,k_y,k_z)P_{\rm
  EM}(k_z,x,y)dk_z
\label{eq:VPS}
\ee
and
\be
\langle \sigma^2(x,y)\rangle=\int P_{\rm 3D}(k_x,k_y,k_z)(1-P_{\rm EM}(k_z,x,y))dk_x
dk_y dk_z,
\label{eq:SPS}
\ee
where $\langle \rangle$ is the ensemble averaging over a number of
realizations, $\langle |\hat V_{\rm 2D}(k_x, k_y,x,y)|^2\rangle$ is an
expectation value of the 2D PS of the observed projected velocity field
$V(x,y)$, $P_{\rm 3D}$ is the PS of the 
3D velocity field and $P_{\rm EM}$ is the PS of normalized
emissivity distribution along the line of sight. For more details see Appendixes A and B. 

\begin{figure}
\plotone{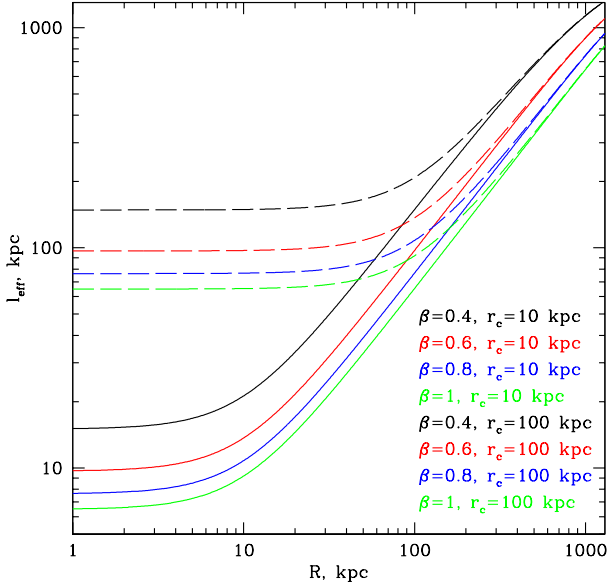}
\caption{Dependence of the effective length along the line of sight $l_{\rm eff}(R)$
  (eq. \ref{eq:leff}) on the projected distance $R$ for different
  $\beta-$models of galaxy clusters.
\label{fig:leff}
}
\end{figure}

In Fig. \ref{fig:sfth} we illustrate the eq. \ref{eq:SPS} for a simple
spherically symmetric $\beta$-model of galaxy cluster with core
radius $r_c=10$ kpc and $\beta=0.6$. 3D velocity PS has a
cored power law model (eq. \ref{eq:psmodel}), i.e. is flat
on $k<k_m=0.005$ kpc$^{-1}$ and has a Kolmogorov slope $\alpha=-11/3$ on
$k>k_m$. Observed velocity dispersion averaged over 100 realizations is shown
with dots, the error bars show the expected uncertainty in one measurement. The right
hand side of eq. \ref{eq:SPS} is shown in red. Minor difference between two
curves at small $R$ is due to finite resolution of simulations.

Once we construct the map of projected velocity $V_{\rm 2D}(x,y)$, one can also
find RMS
velocity of the 2D field  at each distance r from the cluster center as
\be
V_{\rm RMS}(r)=\sqrt{\langle V_{\rm 2D}(x,y)^2\rangle _r-\langle
  V_{\rm 2D}(x,y)_r\rangle^2},
\ee 
where $\langle V_{\rm 2D}(x,y)\rangle_{r}$ denotes mean velocity in ring
at distance r from the center. Below we use the observed velocity dispersion and RMS
of the projected velocity field to constrain the power spectrum.

\section{Structure function and observed velocity
  dispersion}
\label{sec:sf}

Often a structure function of the velocity field is used instead of the power
spectrum, which is defined as
\be
SF(\Delta x)=\langle (V(x+\Delta x)-V(x))^2\rangle,
\ee
where averaging is over a number of pairs of points in space separated
by distance $\Delta x$.
The line-of-sight
velocity dispersion can be linked to the structure
function. Indeed, since the emissivity peaks at the center of the cluster
and declines with the radius, the largest contribution to the total flux
and to the line-of-sight velocity dispersion at distance $R$ from the
center comes from the region, the size of which is $\propto R$.

\begin{figure*}
{\centering \leavevmode \epsfxsize=0.8\columnwidth \epsfbox {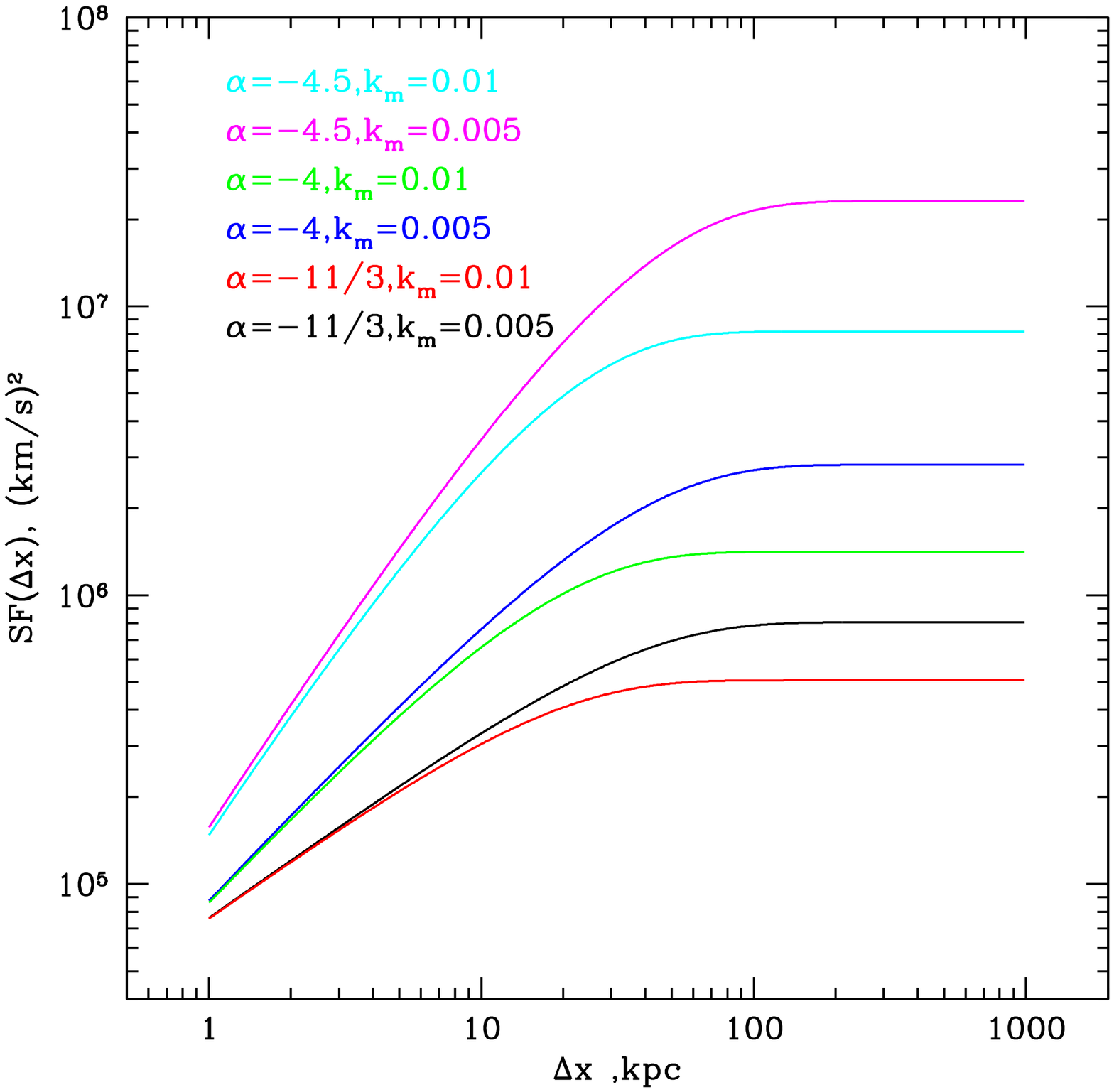}
\centering \leavevmode \epsfxsize=0.8\columnwidth \epsfbox {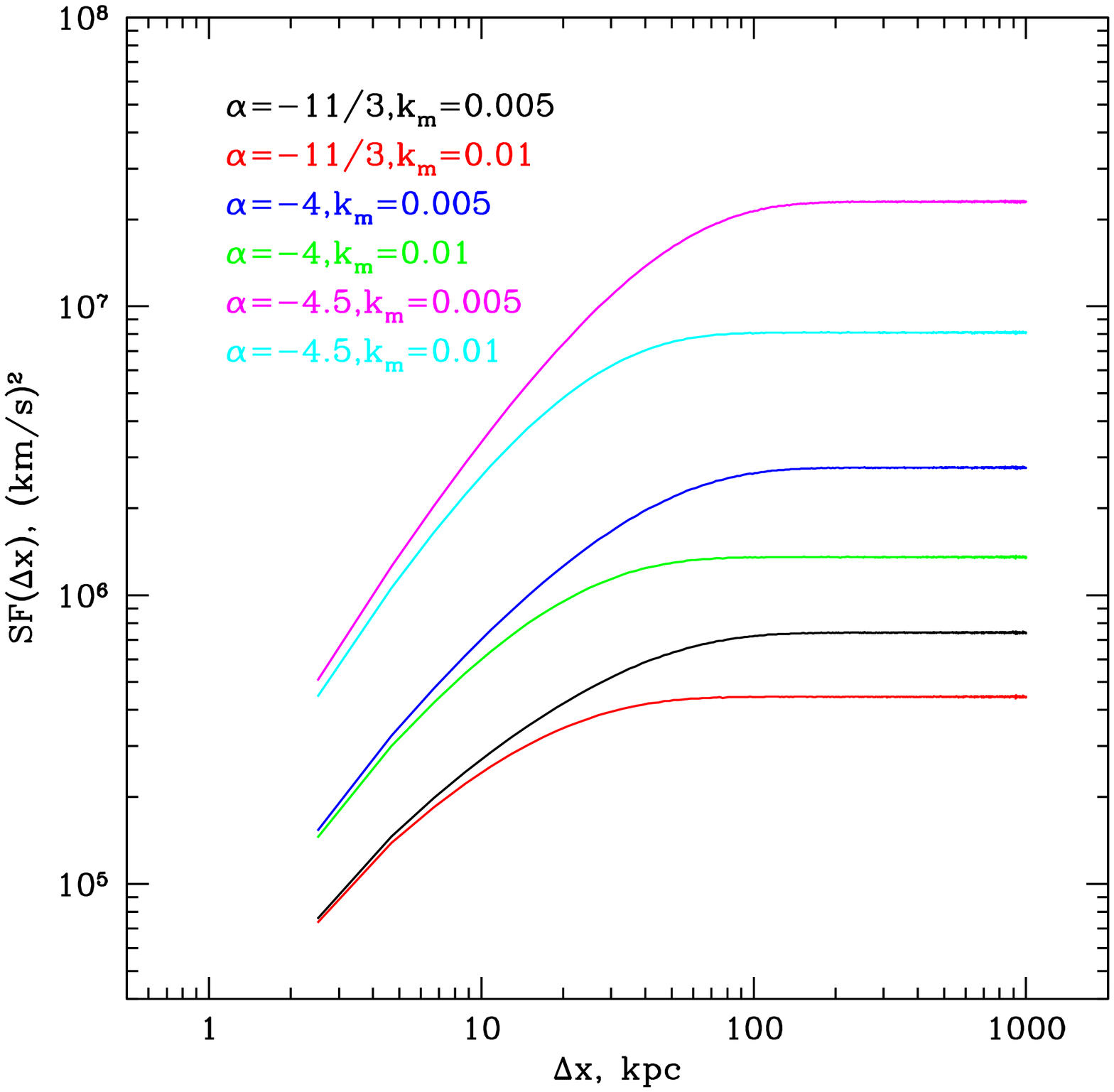}\hfil
\centering \leavevmode \epsfxsize=0.8\columnwidth \epsfbox {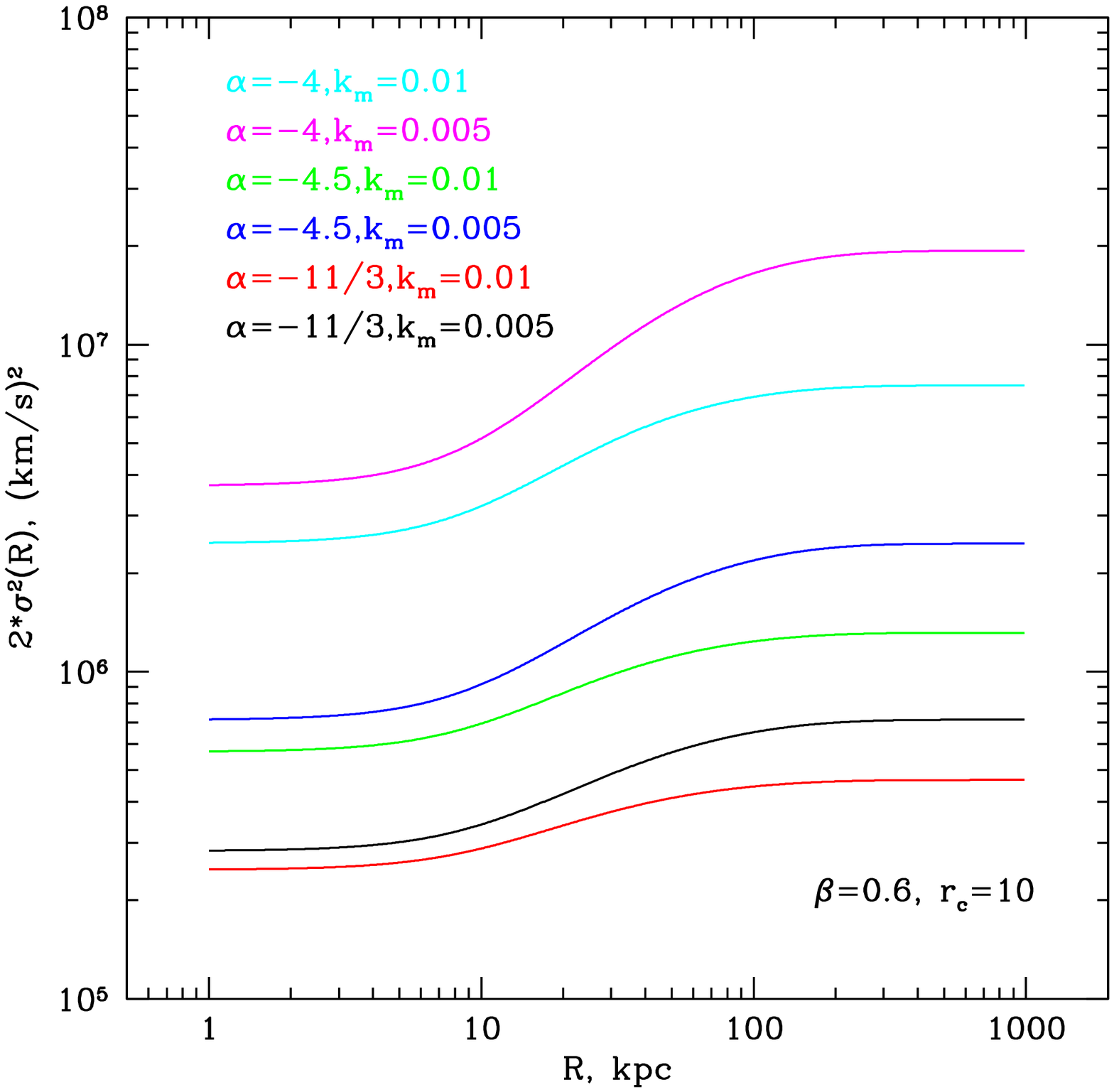}
\centering \leavevmode \epsfxsize=0.8\columnwidth \epsfbox {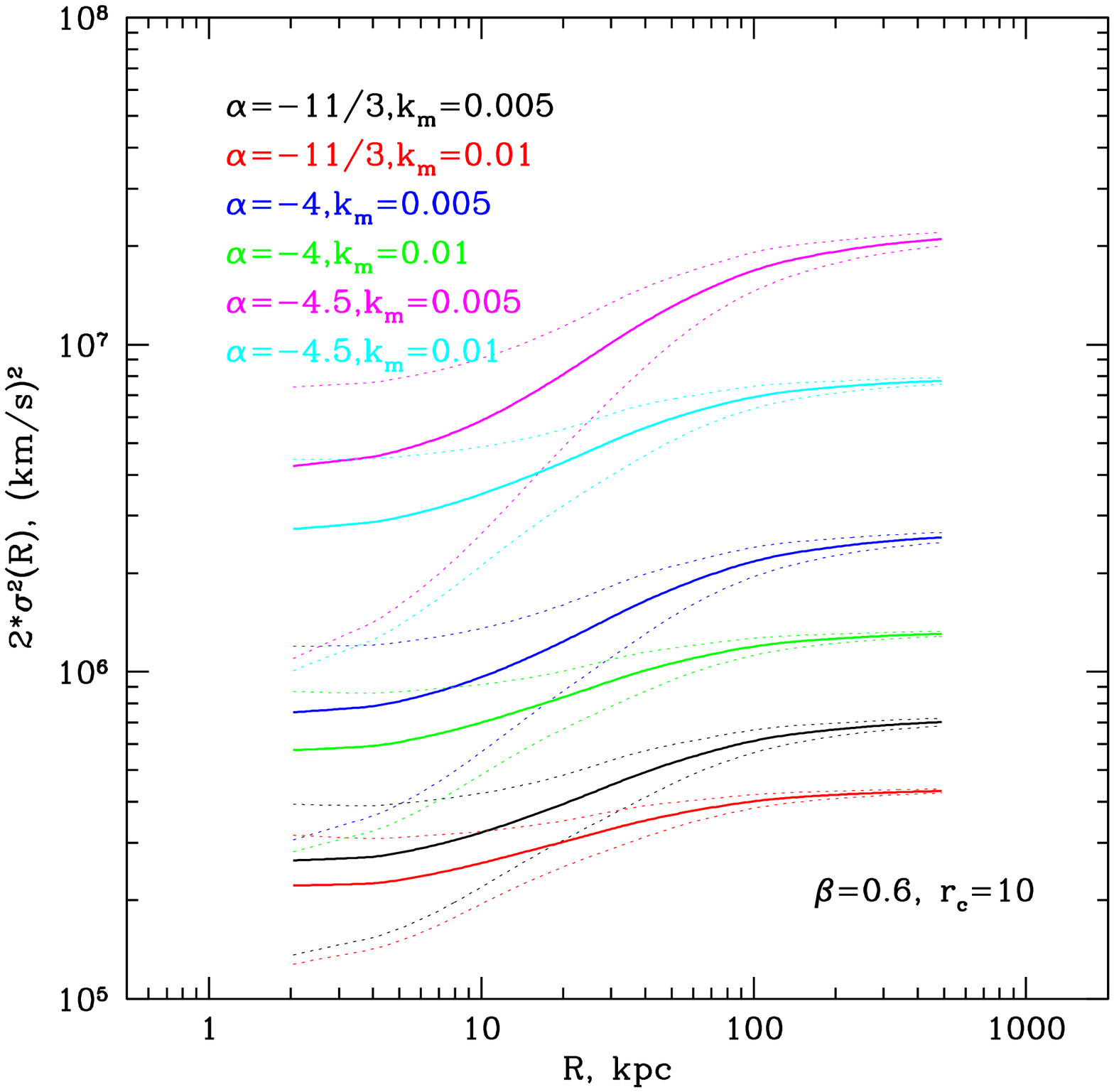}\hfil
\centering \leavevmode \epsfxsize=0.8\columnwidth \epsfbox {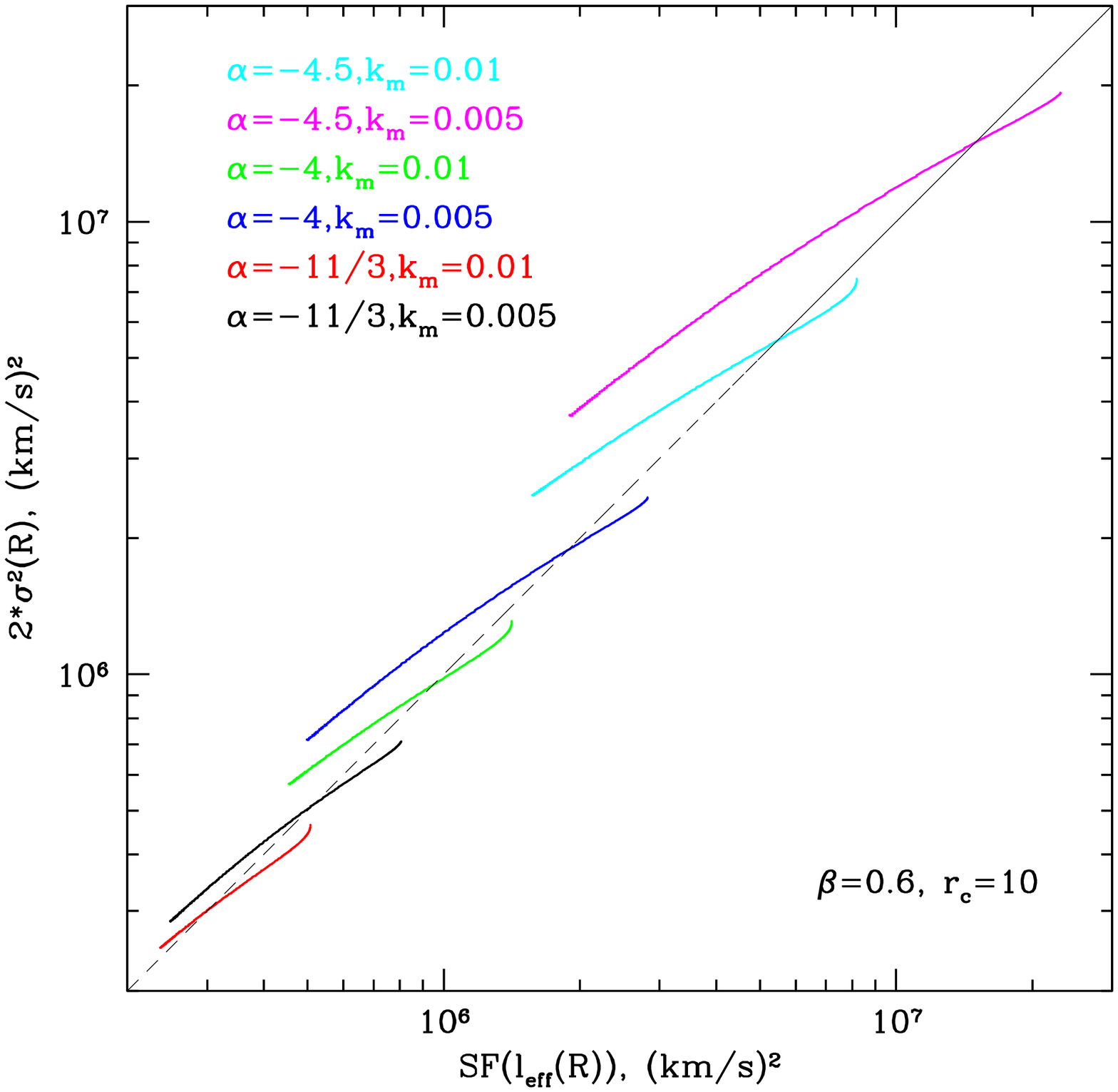}
\centering \leavevmode \epsfxsize=0.8\columnwidth \epsfbox {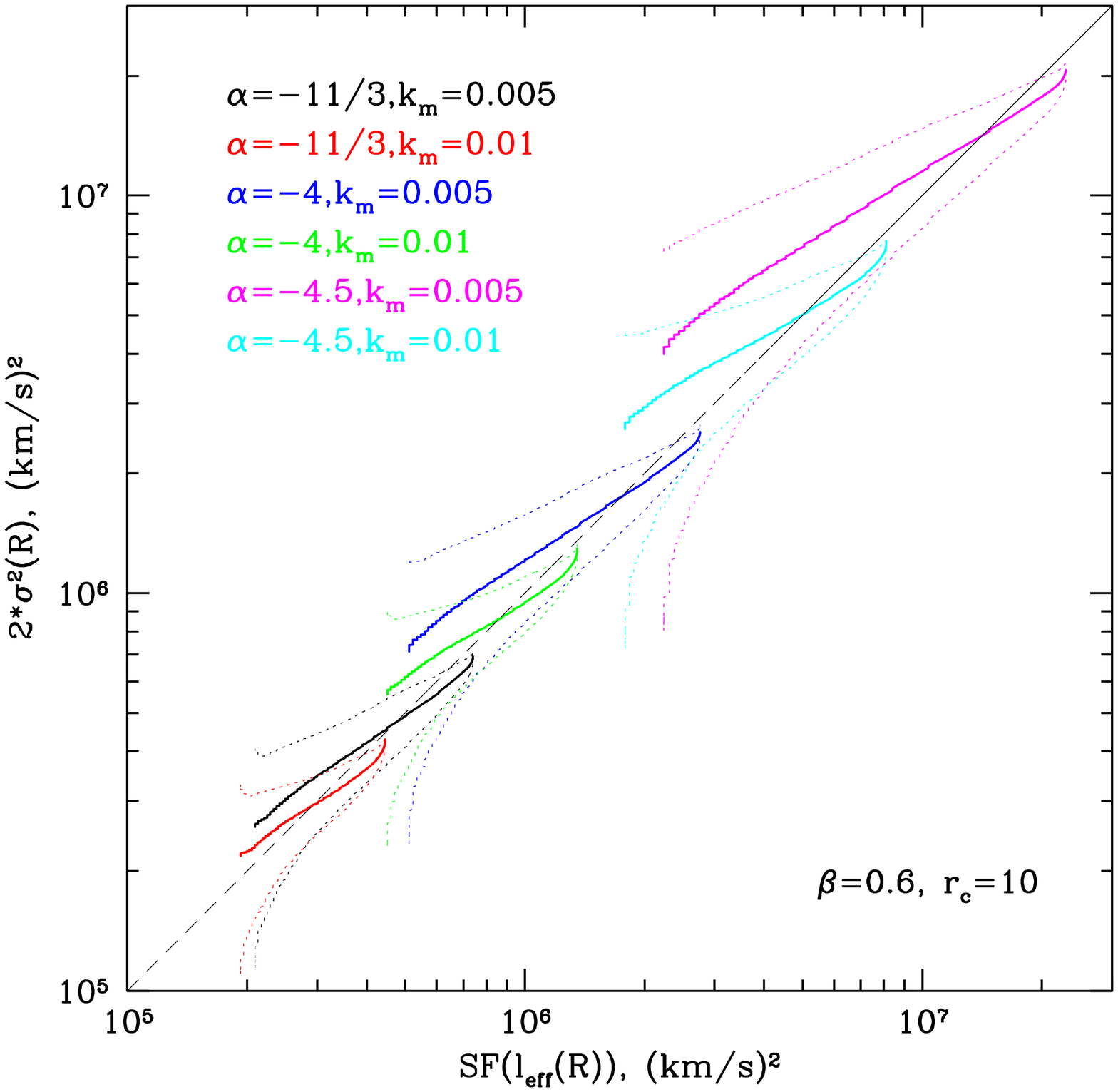}}
\caption{{\bf Left column:} structure function (see analytical expression \ref{eq:sfan}), 
  line-of-sight velocity dispersion multiplied by factor 2
  (eq. \ref{eq:sigma1d} and eq. \ref{eq:weight}) and their relation for different models of cored
  power law power spectrum (eq. \ref{eq:ps3d}). Parameters of the
    power
  spectrum (slope and break wavenumber) are shown in
  the top left corner. Parameters of the $\beta-$ model are given in the
  bottom right corner.\newline
{\bf Right column:} the same as left column, but calculated  by
  averaging over 100 statistical realizations of velocity field. The
  uncertainties in single measurement of the velocity dispersion are
  shown with dotted curves. \newline
 The choice of parameters is discussed in Section 2.
\label{fig:ansim}
}
\end{figure*}

\begin{figure*}
\plottwo{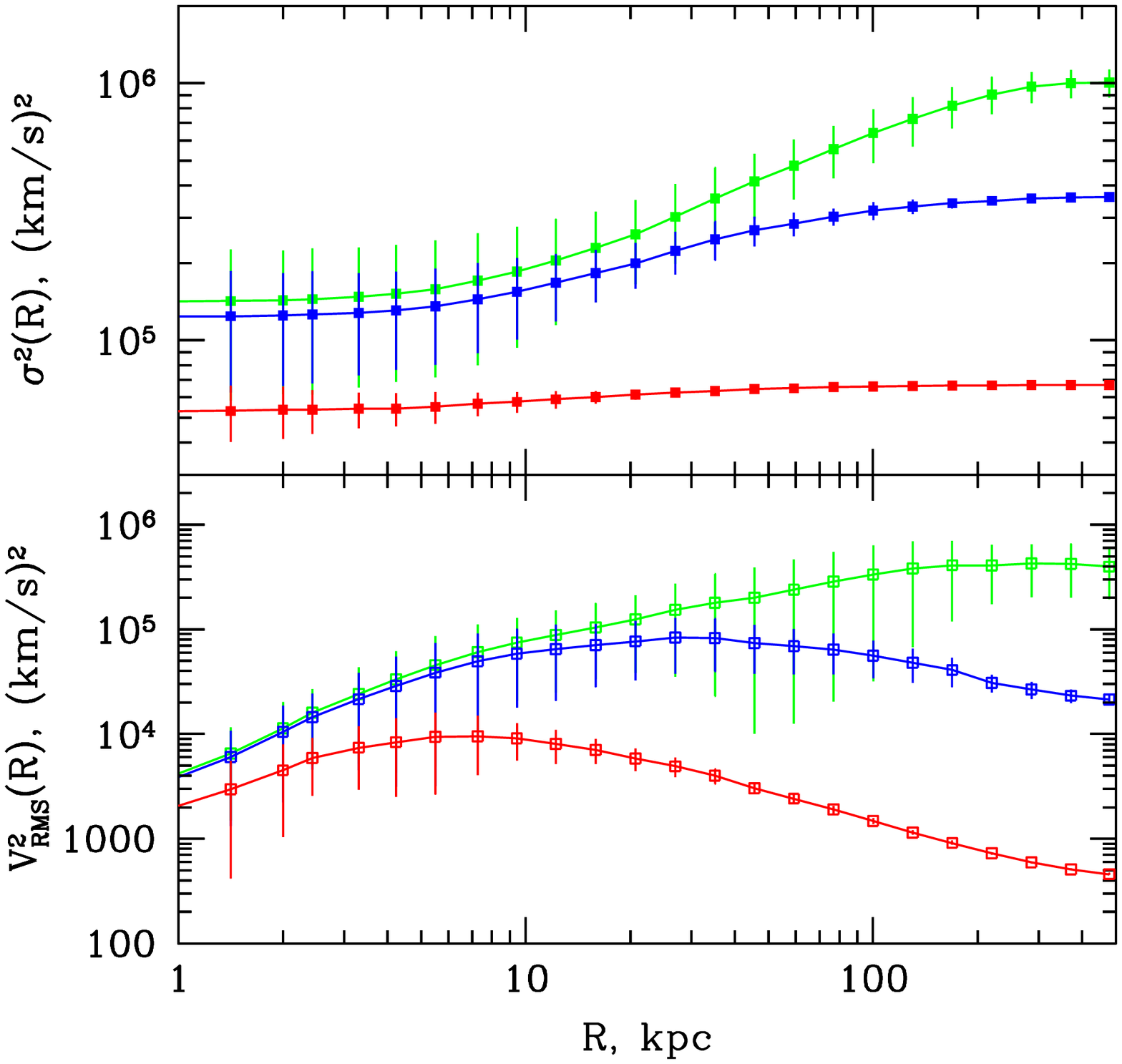}{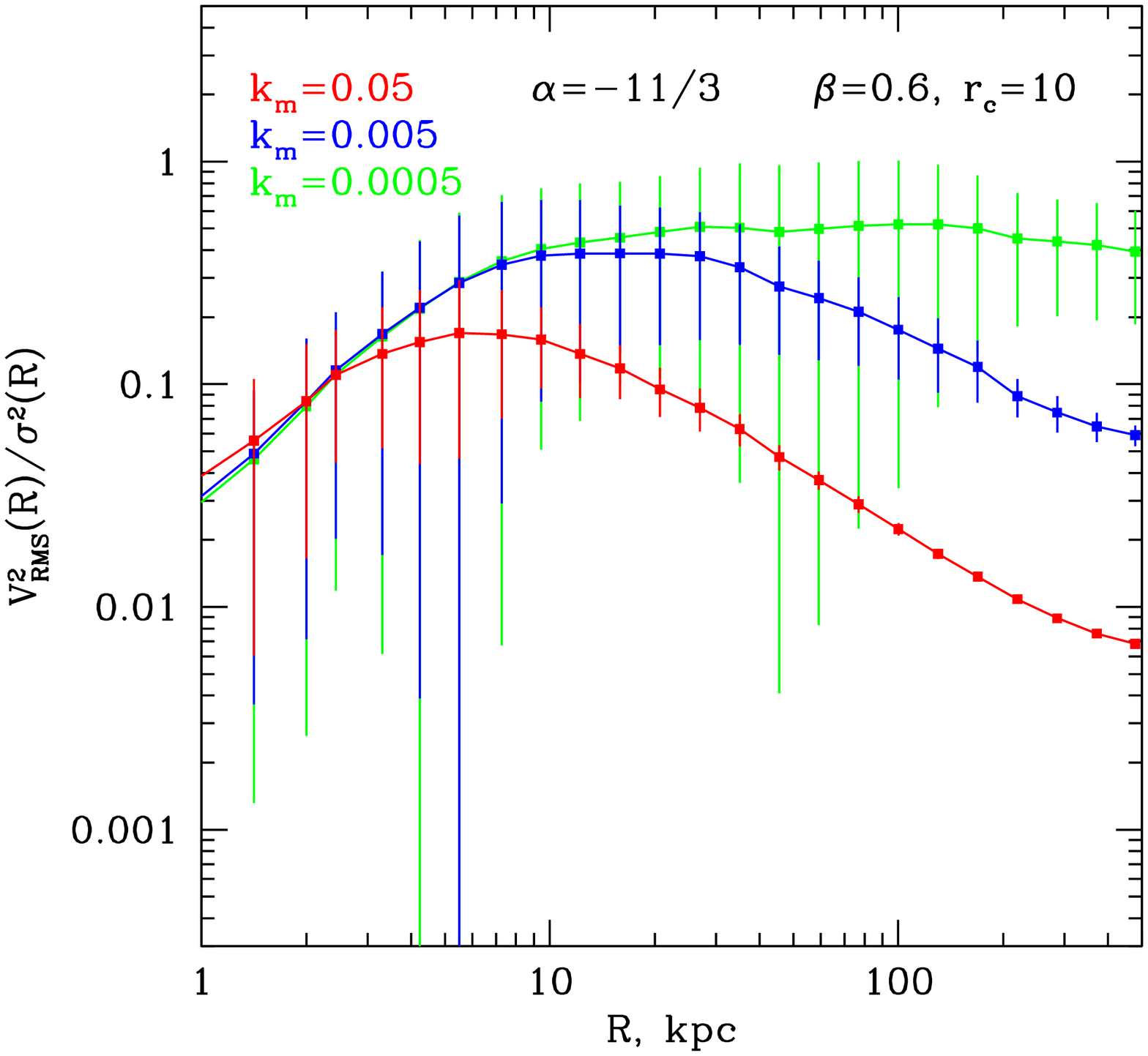}
\caption{ Velocity dispersion and $V_{\rm RMS}$ as functions of
distance from the cluster center with uncertainties due to stochastic nature
of the velocity field (left panels) and their
ratio from eq. \ref{eq:frac} (right panel). $V_{\rm RMS}(R)$ and $\sigma(R)$ are averaged
    over a ring at distance $R$ from the cluster center. Errors show
    the characteristic uncertainties in one measurement. The slope of assumed velocity PS is
  -11/3. Break wavenumber $k_m$ of the PS is shown on the right
  panel. Color coding is the same in both panels.
\label{fig:vrms}
}
\end{figure*}

The structure function and the observed velocity
dispersion can be related to the PS (see appendixes C and D):
\be
SF(x)=2\int_{-\infty}^{+\infty}P_{\rm 1D}(k_{\rm z})(1-\cos 2\pi
k_{\rm z} x ){\rm d} k_{\rm z}
\label{eq:SF1d}
\ee
and
\be
\langle \sigma^2(R) \rangle=\int_{-\infty}^{\infty}P_{\rm 1D}(k_{\rm z})(1-P_{\rm EM}(k_{\rm
  z})){\rm d}k_{\rm z},
\label{eq:S1d}
\ee
where $P_{\rm 1D}$ is an expectation value of the 1D velocity PS and $P_{\rm EM}$ is a PS of normalized emissivity
along the line of sight.
Fig. \ref{fig:integr} shows the integrands in eq. \ref{eq:SF1d} and
eq. \ref{eq:S1d} multiplied by $k$ and $2k$ respectively for the
line of sight near the cluster center (black curves) and at a projected
distance $R=800$ kpc from the center (red curves). Extra factor of 2
for $\sigma^2(R)$ is introduced to compensate for the factor of 2 in
front of the expression \ref{eq:SF1d} for the structure function. 
 It is more clear if one considers the limits of
 these equations at
 large $x$ and $R$. When $x \to \infty$ then $\cos 2\pi
k_{\rm z} x$ oscillates with high frequency over relevant interval
of $k$ and mean value of $1-\cos 2\pi k_{\rm z} x$ is $\sim 1$. When
$R\to \infty$ the emissivity distribution is very broad and
$P_{\rm EM}$ is almost a $\delta$-function. Therefore,
\be
\lim_{x\to \infty}(SF(x))=2\int_{-\infty}^{\infty}P_{\rm 1D}(k_{\rm z})dk_z
\ee
and
\be
\lim_{R\to \infty}(\sigma^2(R))=\int_{-\infty}^{\infty}P_{\rm 1D}(k_{\rm z})dk_z.
\ee
From Fig. \ref{fig:integr} it is clear that the integrands in eq. \ref{eq:SF1d} and eq. \ref{eq:S1d}
are very similar, suggesting that observed $2\sigma^2(R)$ should
correlate well with the structure function.

The structure function and the velocity dispersion (eq. \ref{eq:sfan} and \ref{eq:sigma1d}
respectively) are plotted in Fig. \ref{fig:ansim} in left column. We
fixed parameters of the $\beta-$model of the cluster and varied the
slope $\alpha$ and break $k_m$ of the power spectrum model
(eq. \ref{eq:psmodel}). The relation of SF and $2\sigma^2$ is shown in the left
bottom panel in Fig. \ref{fig:ansim}. For a given $R$, $2\sigma^2(R)$
is used for $x-$axis, while the SF is plotted as a function of $l_{\rm eff}(R)$, where
$l_{\rm eff}$ is an effective length along the line of sight, which
provides dominant contribution to the line flux. $l_{\rm eff}$ is found from the condition that
\be
\disp \frac{\int_0^{l_{\rm eff}}n_e^2(\sqrt{R^2+l^2}) dl}{\int_0^\infty n_e^2(\sqrt{R^2+l^2})
  dl}\approx 0.5.
\label{eq:leff}
\ee
Relation between $l_{\rm eff}$ and projected distance depends on the
$\beta-$ model of galaxy cluster as shown in Fig. \ref{fig:leff}

 We then made multiple statistical realizations of the PS for a simple
$\beta-$model of galaxy cluster with $\beta=0.6$ and $r_c=10$ kpc to
estimate the uncertainties. The size of the box is 1 Mpc$^3$ and resolution
is 2 kpc. We assume that the 3D PS of the velocity field has a
cored power law model (eq. \ref{eq:ps3d}) with
slope $\alpha$ and break wavenumber (injection scale) at $k_m$. We made
100 realizations of a Gaussian field with random phases and
Gaussian-distributed amplitudes in Fourier space. Taking inverse
Fourier transform, we calculated one component of the 3D
velocity field (component along the line of sight) in the
cluster. Structure function and the line-of-sight velocity dispersion
are evaluated using resulting velocity field. Right column in Fig. \ref{fig:ansim} shows velocity
dispersion along the line of sight and structure function averaged over 100
realizations.
 The expected uncertainty in single measurement of
the velocity dispersion is shown with dotted curves. One can see that the
overall shape and normalization of SF and $2\sigma^2$ are the same as
predicted from analytical expressions (left column in
Fig. \ref{fig:integr}), however, there are minor differences
(especially at small $R$) due to limited resolution of
simulations. Relation between $2\sigma^2$ and SF is in a good
agreement with expectation relation, however the uncertainty in
measured velocity dispersion (due to stochastic nature of the velocity
field) is significant (see Section 6).

\section{Length scales of motions and observed RMS velocity of projected velocity field}

Let us now consider RMS of the projected velocity field. During observations one gets
spectra from a region, minimum size of which is set by angular
resolution and/or the sensitivity of the instrument. RMS velocity at certain position (x,y) for random realizations of the velocity field is defined as (see Appendix E)
\be
\langle V^2_{\rm RMS}\rangle=\int P_{\rm 3D}(k_x,k_y,k_z)P_{\rm EM}(k_z)(1-P_{\rm SH}(k_x,k_y))d^3k,
\ee
where $P_{\rm SH}(k_x,k_y)$ is a PS of a mask, where the mask is
defined as zero outside and unity inside the region, from which the spectrum is extracted
(see Appendix E for details).  Velocity dispersion (i.e. the line broadening) measured from the
same region is
\be
\langle\sigma^2\rangle=\int P_{\rm 3D}(k_x,k_y,k_z)(1-P_{\rm
  EM}(k_z)P_{\rm SH}(k_x,k_y))d^3k.
\ee
Therefore the ratio $\langle V^2_{\rm
  RMS}\rangle/\langle\sigma^2\rangle$ is
\be
\frac{\langle V^2_{\rm RMS}\rangle}{\langle \sigma^2\rangle}=\frac{\int P_{\rm
    3D}(k_x,k_y,k_z)P_{\rm EM}(k_z)(1-P_{\rm SH}(k_x,k_y))d^3k}{\int
  P_{\rm 3D}(k_x,k_y,k_z)(1-P_{\rm EM}(k_z)P_{\rm SH}(k_x,k_y))d^3k},
\label{eq:frac}
\ee
which can be used as an additional proxy of the length scales of gas motions.  This
ratio is mostly sensitive to the break of the cored
power law model of the PS $k_m$.  Basically at a given
  line of sight, which is characterized by an effective length $l_{eff}$ the small scale motions (i.e. $k>1/l_{eff}$) are mostly contributing to line broadening, while larger scale motions predominantly contribute to the RMS of the projected velocity field.

 Fig. \ref{fig:vrms} shows the ratio  $\langle V^2_{\rm
  RMS}\rangle/\langle\sigma^2\rangle$  and  its uncertainty
 calculated for different values of parameter
$k_m$.  $V_{\rm RMS}$ and $\sigma$ are averaged over
rings at distance $R$ from the cluster center.  Clearly, if $k_m$ is
large then all motions are on small scales and
 the ratio is small. The larger is the $k_m$, the more
power is at large scales and the larger is the $V_{\rm RMS}$. The larger
is the injection scale, the stronger is the increase of the ratio
with distance and the more prominent is the peak (see
Fig. \ref{fig:vrms}). Here we assume that the full map of projected
velocity field is available. Clearly, the uncertainties will increase
if the data are available for several lines of sight, rather than for
the full map.

Looking at eq. \ref{eq:frac} it is easy to predict behavior of the
ratio on small and large projected distances $R$. Let us specify the shape of area, namely
assume that we measure velocity in circles with radius $R$ around the cluster
center. When $R\to
0$ then the region is very small and $P_{\rm SH}(k_x,k_y)\to 1$ over
broad range of wavenumbers $k< 1/R$ and
the ratio $V^2_{\rm RMS}/\sigma^2\to 0$. At large $R$ $P_{\rm
  SH}(k_x,k_y)\to 0$ and $P_{\rm EM}(k_z)\to 0$ (since $n^2_e(z)$
distribution along the line of sight on large $R$ is broad),
eq. \ref{eq:frac} becomes
\be
\frac{\langle V^2_{\rm RMS}\rangle (R)}{\langle \sigma^2\rangle (R)}=\frac{\int P_{\rm
    3D}(k_x,k_y,k_z)P_{\rm EM}(k_z)d^3k}{\int P_{\rm 3D}d^3k}
\ee
and it is a decreasing function of $R$.

The sensitivity of ratio $V^2_{\rm RMS}/\sigma^2$ to the slope of the
power spectrum is modest. There are only changes in normalization,
but, the overall shape is the same.

\section{Recovering 3D velocity power spectrum from 2D projected
  velocity field}

\begin{figure}
\plotone{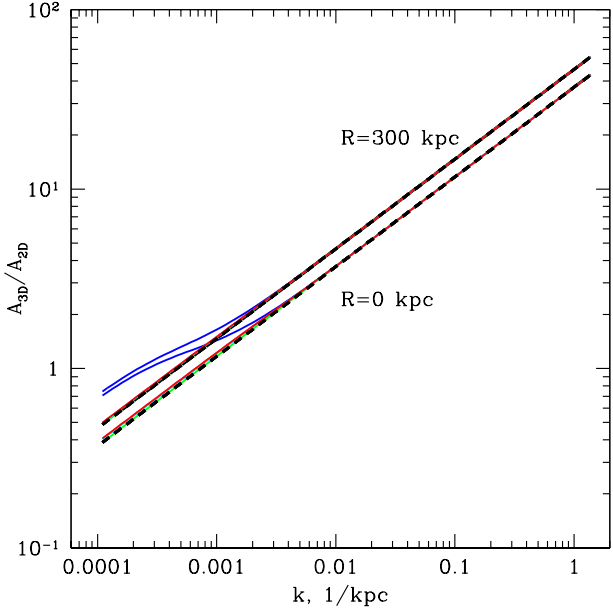}
\caption{ Ratio of 3D and 2D amplitudes as a function of
$k$ (eq. \ref{eq:32}) for a $\beta$-model cluster with
$\beta=0.6$ and $r_c=245$ kpc (Coma-like cluster). A cored power
law model is used for the 3D power spectrum, according to
eq.(\ref{eq:psmodel}). Two projected distances $R$ were used:
$R=0$ and 300 kpc. Solid lines with different colors correspond
to different break wavenumbers in the 3D power spectrum:
$k_m=3\cdot 10^{-4}, 3 \cdot 10^{-3}, 3\cdot 10^{-2}$ kpc$^{-1}$ and different slopes of the spectrum
$\alpha=3.5,~4.0,~4.5$. For a given projected distance the
curves, corresponding to different power spectra model, lay on top
of each other, independently on the slope of the power spectrum, except for the case with very low break
wavenumber $k_m=3\cdot 10^{-4}$ (blue curves). For comparison the
thick dashed lines show the ratio, calculated using simplified
equation (\ref{eq:klarge}). Clearly in the range of scales from Mpc and
below the simple relation between 3D and 2D amplitudes is fully
sufficient to convert the observed 2D power spectrum into the
power spectrum of the 3D velocity field, unless the break of the
power spectrum is at very large scales of few Mpc.
\label{fig:p32km}
}
\end{figure}

\label{sec:3d2d}
Mapping of the projected velocity field $V_{2D}(x,y)$ provides the most
direct way of estimating the 3D velocity field PS. The 2D
and 3D PS are related according to eq.\ref{eq:VPS}, which
we re-write as
\begin{eqnarray}
P_{2D}(k)=\int{P_{3D}(\sqrt{k^2+k_z^2})P_{\rm
  EM}(k_z,x,y)dk_z},
\label{eq:32}
\end{eqnarray}
where $\displaystyle k=\sqrt{k_x^2+k_y^2}$. This equation can be
written as
\begin{eqnarray}
&\nonumber \disp P_{2D}(k)=\int\limits_0^{1/l_{\rm eff}}{P_{3D}(\sqrt{k^2+k_z^2})P_{\rm
  EM}(k_z,x,y)dk_z}+&\\&\disp +\int\limits_{1/l_{\rm eff}}^\infty{P_{3D}(\sqrt{k^2+k_z^2})P_{\rm
  EM}(k_z,x,y)dk_z}&.
\end{eqnarray}
Contribution of the second term to the integral is small since $P_{\rm
  EM}(k_z,x,y)\to 0$ on $k\gg 1/l_{\rm eff}$. In the limit of $k\gg 1/l_{eff}$ (at a given projected distance) the expression reduces to
\begin{eqnarray}
P_{2D}(k)\approx P_{3D}(k)\int{P_{\rm
  EM}(k_z,x,y)dk_z},
\label{eq:klarge}
\end{eqnarray}
i.e. 2D PS is
essentially equal to the 3D PS of the velocity field
apart from the normalization constant $\int{P_{\rm
EM}(k_z,x,y)dk_z}$, which is easily measured for a cluster. We show
below that this simple relation (\ref{eq:klarge}) provides an excellent approximation
for full expression (\ref{eq:32}) for a Coma-like clusters with flat surface brightness
core. For peaked clusters (cool core) $P_{\rm 3D}/P_{\rm 2D}$
  depends on projected distance from the cluster center since $l_{\rm
    eff}$ changes significantly with distance.

It is convenient to use characteristic scale-dependent amplitudes of
the velocity field variations, rather than the PS. The
amplitude for 3D and 2D spectra are defined as
\begin{eqnarray}
A_{3D}(k)= \sqrt{P_{3D}(k)4\pi k^3} \\
A_{2D}(k)= \sqrt{P_{2D}(k)2\pi k^2}
\label{eq:a32}
\end{eqnarray}
In these notations the relation \ref{eq:klarge} between PS transforms to
\begin{eqnarray}
A_{2D}(k)= A_{3D}(k)\sqrt{\frac{1}{2}\frac{\int{P_{\rm
  EM}(k_z,x,y)dk_z}}{k}}.
\label{eq:a3to2}
\end{eqnarray}
Integral $\int P_{\rm EM}(k_z,x,y)dk_z$ can be estimated as
  $\disp\frac{1}{l_{\rm eff}}$, since the largest contribution is on 
$k<1/l_{\rm eff}$. Eq. \ref{eq:a3to2} becomes
\be
A_{2D}(k)= A_{3D}(k)\sqrt{\frac{1}{2}\frac{1}{l_{\rm
        eff}k}}=A_{3D}(k)\sqrt{\frac{1}{2}\frac{1}{N_{\rm
          edd}}}.
\ee
The essence of this relation is that the amplitude of the 3D velocity
fluctuations is attenuated in the 2D projected velocity field by a
factor of order $\displaystyle\sqrt{ \frac{1}{N_{edd}}}$, where $N_{edd}$ is
the number of independent eddies which fit into effective length along
the line of sight.

We illustrate the above relation for the case of the Coma cluster. The
density distribution in the Coma can be characterized by a $\beta$-model
with $\beta=0.6$ and co-radius $r_c=245$ kpc. In Fig. \ref{fig:p32km} we
plot the ratio $A_{3D}(k)/A_{2D}(k)$ evaluated using equations (\ref{eq:klarge}) and (\ref{eq:32})
 for a number of 3D PS models calculated at two
projected distances from the Coma center. One can see that on spatial
scales of less than 1 Mpc the equation (\ref{eq:a3to2}) is fully
sufficient. The variations of the relation for different projection
distances (projected distances from 0 to 300 kpc were used) affect
only the normalization of the relation and can easily be accounted
for.

With ASTRO-H the 2D velocity field in the Coma can be mapped with the
$\sim 1.7'$ resolution, which corresponds to $\sim 46$ kpc. Mapping
$450\times 450$ kpc central region of the Coma would require about 36
pointings. For practical reasons it may be more feasible to make a
sparse map (e.g. two perpendicular stripes) to evaluated $A_{2D}$ \citep[
e.g. computing correlation function or using a method described in] []{Are11}

\section{Discussion}

\subsection{Limiting cases of small and large scale motions}

 Measuring characteristic amplitude of mean velocity ($V_{\rm RMS}$)
and velocity dispersion ($\sigma$) we can distinguish whether the
turbulence is dominated by small or large scale
motions. Clearly, the motions on scales much smaller than the effective
length along the line of sight near the cluster center can only
contribute to the line broadening. This sets the characteristic value
of the lowest spatial scale which can be measured. The largest
measurable scale is set by the maximum distance $R_{\rm max}$ from the cluster
center where the line parameters can be accurately measured without
prohibitively long exposure time. Thus the range of scales $l$, which can
be probed with these measurements is $l_{\rm eff}(R=0)<l<R_{\rm max}$.
The crucial issue in
measurements is ``sample variance'' of measured quantities caused by
stochastic nature of turbulence. We can expect two limiting
cases (see Fig. \ref{fig:vrms}). \newline
{\bf A:} {\it Small scale motions}\\
 In case of small scale motions $l\ll l_{\rm eff}(R=0)$ (i.e. $k_m \gg
l_{\rm eff}(R=0)$ one expects $\sigma(R)$ to be independent of radius
and $V_{\rm RMS}(R)\ll\sigma(R)$. $\sigma(R)$ is expected to have low
sample variance and can be measured
accurately even for a single line of sight, provided sufficient
exposure time.  Measurements of $V_{\rm RMS}(R)$ are strongly
affected by sample variance and depend on the geometry of the measured
map of the projected velocity dispersion $V(x,y)$. If $V(x,y)$ is
measured only at two positions, then the
uncertainty in $V_{\rm RMS}(R)$ is of the order of its value and the
ratio $\sigma(R)/V_{\rm RMS}(R)$ gives only low limit on $k_m$. 
We note that in this limit of small scale motions the assumption of a
uniform and homogeneous Gaussian field can be relaxed and measured
values of line broadening $\sigma$ simply reflect the total variance
of the velocity along the given line of sight, while the possibility
of determining the spatial scales of motions are limited. Variations
of $\sigma$ with radius will simply reflect the change of the
characteristic velocity amplitude. \newline
{\bf B:} {\it Large scale motions}\\
 In case if most of turbulent energy is associated with large scales (i.e. $k_m \ll
l_{\rm eff}(R=0)$, $\sigma(R)$ is expected to increase with $R$
and $V_{\rm RMS}(R)\sim \sigma(R)$. In this case sample variance
affects both $\sigma(R)$ and $V_{\rm RMS}(R)$. Mapping the whole area
(as opposed to measurements at few positions) would help to reduce
the sample variance. Knowing the shape of $\sigma(R)$ and estimates of
$k_m$ from $\sigma(R)/V_{\rm RMS}(R)$ we can constrain the slope of power
spectrum.

\subsection{Effect of thermal broadening}
\label{sec:dislines}

\begin{figure}
\plotone{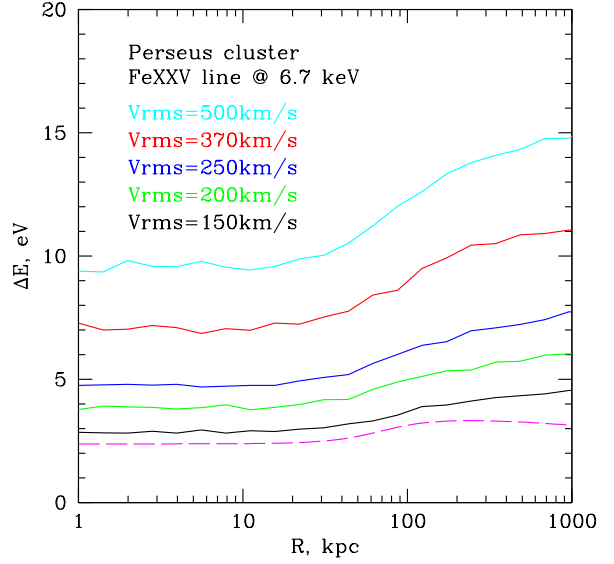}
\caption{Width of the line of He-like iron line at 6.7 keV in Perseus
  cluster, assuming different 1D RMS velocities. FWHM is related with
  the width as $FWHM=1.66\Delta E$. Solid curves show
  broadening due to gas motions only, dashed curve shows the thermal
  broadening. Here we assumed the cored power law 3D power spectrum
  with $\alpha=-11/3$ and $k_m=0.001$ kpc$^{-1}$.
\label{fig:perde}
}
\end{figure}

 Measuring velocity dispersion one should account for line
  broadening due to thermal motions of ions. Thermal broadening should
be subtracted from measured width of line.  The broadening of the line $\Delta
E=\sqrt{2}\sigma$, where $\sigma$ is measured Gaussian width of the
line, is defined as
\be
\Delta E=\frac{E_0}{c}\sqrt{2(\sigma^2_{therm}+\sigma^2_{turb})},
\label{eq:deltaE}
\ee where $\sigma_{turb}$ is the width due to turbulent motions and
$\displaystyle \sigma_{therm}=\sqrt{\frac{kT}{A m_p}}$ is the thermal
broadening for ions with atomic weight $A$. The FHWM of the
corresponding line is ${\rm FWHM}=2\sqrt{\ln2}\Delta E\approx 1.66 \Delta E$.  Observed emission
  lines in the X-ray spectra
of galaxy clusters correspond to heavy elements such as Fe, Ca,
S. Because
of large value of $A$ the contribution of thermal broadening is 
small even for modest amplitudes of turbulent velocities. Indeed, the atomic
  weight for iron is 56 and the thermal width of the iron line is
  $\Delta E\sim 3$ eV (${\rm FWHM}\sim 5$ eV) if one assumes the typical temperature of clusters $\sim
  5$ keV. At the same the gas motion with the sound speed would causes
the shift of the line energy by $\sim$40 eV. 
As an example, we calculated the expected broadening of the He-like
iron line at 6.7 keV for the Perseus cluster, assuming Kolmogorov-like
PS of the velocity field with $k_m=0.001$ kpc$^{-1}$ and varying
total RMS of the velocity field in one dimension. The
model of the Perseus cluster was taken from \citet{Chu04} and modified
at large distances according to Suzaku observations at the edge of the
cluster \citep{Sim11}, i.e. the electron number density is
\be
N_e(r)=\frac{4.68\cdot 10^{-2}}{\left[1+\left(\disp\frac{r}{56}\right)^2\right]^{1.8}}+\frac{4.86\cdot 10^{-3}}{\left[1+\left(\disp\frac{r}{194}\right)^2\right]^{0.87}}
\ee
and the temperature profile is
\be
T(r)=7\frac{1+\left(\disp\frac{r}{69}\right)^3}{2.3+\left(\disp\frac{r}{69}\right)^3}\times\left(1+\disp\frac{r}{5000}\right)^{-1}.
\ee
Abundance of heavy elements is assumed to be constant 0.5 relative to
Solar \citep{And89}. Fig. \ref{fig:perde} shows the calculated width of 6.7 keV line
assuming various 1D RMS velocities. Thermal broadening
is shown with the dashed magenta curve. One can see that thermal
broadening starts to dominate broadening due to motions only if
$V_{RMS}< 150$ km/s. However, the lack of resonant scattering signatures in the
spectrum of the Perseus cluster suggests that the expected velocity is higher than 400
km/s in the center of the Perseus cluster \citep{Chu04}. ${\it Astro-H}$ will have energy resolution of 7 eV at 6.7 keV,
therefore broadening due to gas motions with $V_{\rm
  RMS}\propto$ 400 km/s will be easy to observe
(Fig. \ref{fig:perde}) in the Perseus cluster.

\begin{figure}
\plotone{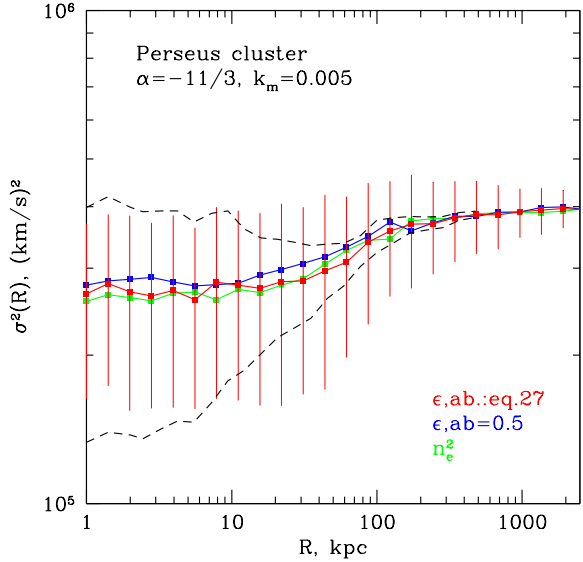}
\caption{ Velocity dispersion in the Perseus cluster as function of
  distance from the center for different emissivity and abundance
  profiles. Parameters of assumed velocity PS are
  shown in the top left corner. Green: simple approximation of
  emissivity $\approx n_e^2$; blue: emissivity at 6.7 keV line and
  flat abundance; red: emissivity at 6.7 keV line and peaked abundance
  abundance (eq. \ref{eq:pab}) in the center (see text for
  details). Black dashed curve shows expected uncertainty in measurements
  of velocity dispersion observed in a ring.
\label{fig:peremis}
}
\end{figure}

\subsection{The effect of radial variations of the
  gas temperature and metallicity}

The analysis described in the previous sections was done assuming
an isothermal $\beta$-model spherically symmetric cluster with emissivity simply
$\propto n_e^2$. Clearly, real clusters are more complicated,
even if we keep the assumption of spherical symmetry. First of all the
gas temperature and metallicity often vary with radius. These
variations will be reflected in the weighting function $\displaystyle
n_e^2(\sqrt{R^2+z^2})$, which relates 3D velocity field and observables.
To verify how strongly these assumptions
affect the results, we calculated velocity dispersion assuming
detailed model of the Perseus
cluster (see above). We assumed constant abundance profile and more
realistic peaked abundance profile, taken from {\it Suzaku} \citep{Sim11}
 and {\it Chandra/XMM} observations 
\be
Z(r)=0.4\frac{2.2+\left(\disp\frac{r}{80}\right)^2}{1+\left(\disp\frac{r}{80}\right)^2}.
\label{eq:pab}
\ee

We calculated the emissivity of the He-like iron line at 6.7 keV and used this
emissivity as a weighting function for the calculation of the expected
projected velocity and 
velocity dispersion. Fig. \ref{fig:peremis} shows the velocity
dispersion along one line of sight calculated for the
most simple model of emissivity and assuming more complicated models
described above. One can see that mean value and uncertainties are
very similar in all cases. Clearly that averaging velocity dispersion
over a ring will decrease the uncertainty in one measurement
(Fig. \ref{fig:peremis}, black
dashed curve).

\subsection{Influence of density fluctuations}
\label{sec:denfl}

The hot gas density in galaxy clusters is strongly peaked towards the
centre. However, besides the main trend, there are density
fluctuations, which could contribute to the observed fluctuations of
the projected velocity. Let
us split the density field in two components 
\be
n=n_0+\delta n,
\label{dne}
\ee
where $n_0$ corresponds to the smooth global profile and $\delta n$ represents
fluctuating part. Neglecting terms of the order $\left (\delta n
\right )^2$, the
emissivity-weighted projected velocity is
\be
V_{\rm 2D}(x,y)=\frac{\int n_0^2(z)V(x,y,z)dz + 2\int n_0(z)\delta n
  V(x,y,z) dz}{\int n_0^2(z)dz}.
\label{Vd}
\ee
The analysis of X-ray surface brightness fluctuations in the Coma
cluster \citep[e.g.][]{Chu12} have shown that the amplitude of the
density fluctuations $\langle \frac{\delta n}{n_0}\rangle$ 
is of the order $\sim 2-10 \%$. Assuming that the same number is
applicable to other clusters,  
the second term in eq. \ref{Vd} is small and the contribution of the
density fluctuation to projected velocity can be neglected.

\subsection{Measurements requirements}

\begin{table}
 \centering
  \caption{ Energy and angular resolutions requirements for future
    X-ray observatories.} 
  \begin{tabular}{lcc}
  \hline
  & Perseus cluster & NGC5813\\
  & FeXXV line & OVIII line\\
\hline 
 FWHM (therm.)& 4.9 eV & 0.32 eV \\
 Shift of line energy (V=c$_s$) & 25.7 eV & 0.91 eV \\
 Angular resolution & 0.5$'$ & 2$''$ \\
 \hline
\label{tabinstr}
\end{tabular}
\end{table} 

To illustrate the most basic requirements for the instruments to measure the ICM
velocity field,  let us consider two examples of a rich cluster and an
individual elliptical galaxy, representing the low temperature end of
the cluster-group-galaxy sequence:\\
\begin{enumerate}
\item Perseus cluster, $T_{mean}=5$ keV, line of FeXXV at 6.7 keV,
  distance 72 Mpc
\item Elliptical galaxy NGC5813, $T_{mean}=0.65$ keV, line of OVIII at
0.654 keV, distance 32.2 Mpc
\end{enumerate}

Table \ref{tabinstr} shows the FWHM of these lines calculated if only
thermal broadening is taken into account, shift of the line energy in the case of gas motions with $V=c_s$ and desirable angular resolution of the instrument.
Future X-ray missions, such as {\it Astro-H} and {\it ATHENA} will have energy
resolution $\sim 5$ eV and $\sim 3$ eV respectively. Such energy
resolution is sufficient to measure broadening of lines in hot systems
like galaxy clusters. Cold systems, like elliptical galaxies, require
even better energy resolution. However, turbulence still can be
measures using the resonant scattering effect \citep[see e.g.][]{Wer09,Zhu11} or by using
grating spectrometers observations. Angular resolution of $\sim 1'$
for nearby galaxy clusters should sufficient to study the most
basic characteristics of the ICM velocity field, while for elliptical
galaxies the resolution at the level of $\sim$ arcsec (comparable with
Chandra resolution) would be needed.

{\it Astro-H} observatory will have energy resolution $\sim 5$ eV,
field-of-view $2.85'$ and angular resolution $1.7'$, which means that it
will be possible to measure shift and broadening of lines  as a
function of projected distance from the center in nearby
clusters. E.g. if one assumes that RMS velocity of gas motions in
Perseus cluster is $\sim 300$ km/s, then $\sim 4\cdot 10^5$ s is enough to
measure profiles of mean velocity and velocity dispersion with
a statistical uncertainty of $\sim 30$ km/s (90\% confidence) in a stripe $\pm$ 200 kpc (7 independent
pointing and 28 independent measurements in $1.4' \times 1.4'$ pixels) centered in the center of the cluster \footnote{The estimates were done using
the current version of {\it Astro-H} response at http://astro-h.isas.jaxa.jp/researchers/sim/response.html}. In order to measure velocity with the
same accuracy at larger distances from the center , e.g. at 500 kpc
and 1 Mpc, one would need $8\cdot 10^5$ s and $4 \cdot 10^6$ s
exposure respectively.

\section{Conclusions}
 Various methods of constraining the velocity power
  spectrum through the  observed shift of line centroid and line broadening are
  discussed. 
\begin{itemize}
\item Changes of the line broadening with
  projected distance reflects the increase of the spread in the
  velocities with distance, closely resembling the behavior of the
  structure function of the velocity field.
\item Another useful quantity
  is the ratio of the characteristic amplitude of the projected
  velocity field to the line broadening. Since the projected velocity
  field mainly depends on large scale motions, while the line
  broadening is more sensitive to small scale motions, this ratio is a
  useful diagnostics of the shape of the 3D velocity field power
  spectrum. 
\item Projected 2D velocity field power spectrum can be easily
  converted into 3D power spectrum. This conversion is especially
  simple for cluster with an extended flat core in the surface
  brightness (like Coma cluster). 
\end{itemize}
Analytical expressions are derived for a $\beta$-model
clusters, assuming homogeneous isotropic Gaussian 3D velocity
field. The importance of the sample variance, caused by the stochastic
nature of the turbulence, for the observables is
emphasized.

\section{Acknowledgements} 
IZ, EC and AK would like to thank  Kavli Institute for Theoretical Physics (KITP) in Santa
Barbara for hospitality during workshop "Galaxy clusters: crossroads of
astrophysics and cosmology" in March-April 2011, where part of the work
presented here was carried out. This research was supported in part by the National Science Foundation under Grant No. NSF PHY05-51164.
IZ would like to thank the International Max Planck Research School
on Astrophysics (IMPRS) in Garching.

\appendix

\section[]{3D velocity power spectrum and projected velocity
  field}
Let us assume that the line-of-sight component of the 3D velocity
field $V_{\rm 3D}(x,y,z)$ is described by a Gaussian (isotropic
and homogeneous) random field. We assume that the centroid shift and the width of lines contain most
essential information on the velocity field. Here and below we
adopt the relation $k=1/r$ without factor $2\pi$ (see Section 2 for details).

Projected 2D velocity along the line of sight (observed centroid shift
of the emission line) in $z$ direction is
\be
V_{\rm 2D}(x,y)=\frac{\int V_{\rm 3D}(x,y,z)n^2_{\rm e}(x,y,z) dz}{\int
  n^2_{\rm e}(x,y,z) dz},
\ee
where $V_{\rm 3D}$ is $z$ component of the 3D velocity field and
$n_{\rm e}$ is the electron
number density. Denoting normalized emissivity
along the line of sight at a certain position with coordinates (x,y) as $\epsilon(z)=n^2_{\rm e}(x,y,z)/\int n^2_{\rm
  e}(x,y,z) dz$ the previous relation can be re-written as
\be
V_{\rm 2D}(x,y)=\int V_{\rm 3D}(x,y,z)\epsilon (z) dz.
\label{eq:v2d}
\ee
Applying the convolution theorem one can find the Fourier transform of
$V_{\rm 3D}(x,y,z)\epsilon (z)$ as
\be
\int \hat V_{\rm 3D}(k_x,k_y,k_{z_1}) \hat \epsilon (k_z-k_{z_1})
dk_{z_1},
\label{ftv3d}
\ee
where $\hat V_{\rm 3D}$ and $\hat \epsilon$ are Fourier transforms of the
3D velocity field and normalized emissivity respectively.
The projection-slice theorem states that
\be
\hat f_{\rm 2D}(k_x,k_y)=\hat f_{\rm 3D}(k_x,k_y,0).
\label{prslth}
\ee
Accounting for \ref{ftv3d} and \ref{prslth} we can write the Fourier
transform of 2D velocity field as
\be
\hat V_{\rm 2D}(k_x,k_y)=\int \hat V_{\rm 3D}(k_x,k_y,k_{z_1})\hat
\epsilon ^*(k_{z_1}) dk_{z_1},
\ee
where * denotes conjugation.

Averaging over a number of realization we find the power spectrum
of the projected mean velocity 
\be
\langle|\hat V_{\rm 2D}(k_x, k_y)|^2\rangle=\langle|\int \hat V_{\rm 3D}(k_x,k_y,k_{z_1})\hat
\epsilon^*(k_{z_1})dk_{z_1}|^2\rangle.
\ee
The right part of the equation above can be re-written as
\beq
&\nonumber \disp \int \langle\hat V_{\rm
  3D}(k_x,k_y,k_{z_1})\hat V^*_{\rm
  3D}(k_x,k_y,k_{z_2})\rangle\times&\\ &\disp \times \hat \epsilon
^*(k_{z_1})\hat \epsilon (k_{z_2})dk_{z_1} dk_{z_2}.&
\eeq
Since phases are random, all cross terms after averaging over a number
of realizations will give 0 if $k_{z_1} \ne k_{z_2}$. Therefore,
\be
\langle|\hat V_{\rm 2D}(k_x, k_y)|^2\rangle=\int |\hat V_{\rm
  3D}(k_x,k_y,k_{z_1})|^2 |\hat \epsilon (k_{z_1})|^2 dk_{z_1}.
\ee
Denoting power spectra of the 3D velocity field and normalized
emissivity as $P_{\rm 3D}$ and $P_{\rm EM}$ respectively the final
expression is
\be
\langle|\hat V_{\rm 2D}(k_x, k_y)|^2\rangle=\int P_{\rm 3D}(k_z,k_y,k_z)P_{\rm
  EM}(k_z)dk_z.
\label{meanV}
\ee

\section[]{3D velocity power spectrum and projected velocity dispersion}

Projected mean velocity dispersion for the line of sight with
coordinates (x,y) averaged over a number of
realization is defined as

\beq
&\nonumber \disp \langle\sigma^2(x,y)\rangle=\langle\int V^2_{\rm 3D}(x,y,z)\epsilon (z)dz\rangle
-&\\& \disp -\langle \left ( \int V_{\rm
  3D}(x,y,z)\epsilon (z)dz \right) ^2\rangle.&
\eeq
It can be re-written as
\beq
&\nonumber \disp 
\langle\sigma^2(x,y)\rangle=\int\langle V^2_{\rm 3D}(x,y,z)\rangle\epsilon (z)dz
-&\\&\disp -\langle V^2_{\rm 2D}(x,y)\rangle.&
\eeq
Expanding $V_{\rm 3D}$ and $V_{\rm 2D}$ through the Fourier
series, averaging over realizations and keeping non-zero cross
terms will give
\beq
&\nonumber \disp \langle\sigma^2(x,y)\rangle=\int |\hat V_{\rm 3D}(k_x,k_y,k_z)|^2dk_x dk_y dk_z\int
\epsilon (z) dz -&\\&\disp -\int |\hat V_{\rm 2D}(k_x,k_y)|^2dk_x dk_y.&
\eeq
Since emissivity along the line of sight is normalized so that $\int
\epsilon (z)dz=1$ and accounting for eq.\ref{meanV} the final expression for
projected velocity dispersion is
\be
\langle\sigma^2(x,y)\rangle=\int P_{\rm 3D}(k_x,k_y,k_z)(1-P_{\rm EM}(k_z))dk_x
dk_y dk_z.
\label{eq:sigPS}
\ee

\section[]{Relation between structure function and cored power law 3D
  power spectrum}

Let assume 3D isotropic and homogeneous power spectrum (PS) of the velocity field is described as

\be
P_{\rm 3D}(k_{\rm x},k_{\rm y},k_{\rm z})=\frac{B}{\left(1+\disp\frac{k_x^2+k_y^2+k_z^2}{k_{\rm
      m}^2}\right)^{\alpha/2}},
\label{eq:ps3d}
\ee
where $k_{\rm m}$ is a wavenumber where $\beta$ model has a break
(e.g. an injection scale in turbulence model),
$\alpha$ is a slope of PS on $k>k_{\rm m}$ and $B$ is PS normalization, which is defined so
that the characteristic amplitude $A$ of velocity fluctuations at $k_{\rm ref}$ is fixed, i.e.

\be
B=\frac{A^2}{4\pi k_{\rm ref}^3 P_{\rm 3D}(k_{\rm ref})}.
\ee 

Integrating \ref{eq:ps3d} over $dk_xdk_y=2\pi kdk$, one can find 1D PS 

\be
P_{\rm 1D}(k_{\rm z})=\frac{\disp 2\pi B k_{\rm m}^2\left(1+\frac{k_{\rm
      z}^2}{k_{\rm m}^2}\right)^{-\alpha/2+1}}{\alpha-2}.
\label{ps1d}
\ee
Structure function (SF) is related to 1D PS by the transformation
\citep{Ryt88}
\be
SF(x)=2\int_{-\infty}^{+\infty}P_{\rm 1D}(k_{\rm z})(1-\cos 2\pi
k_{\rm z} x ){\rm d} k_{\rm z}.
\label{eq:sfps}
\ee
Substituting \ref{ps1d} to \ref{eq:sfps} and assuming $\alpha >3$ yields
\be
SF(x)=\frac{\disp 4Bk_{\rm
    m}^3\left(\disp \pi^\frac{3}{2}\Gamma(\xi)-2\pi^\frac{\alpha}{2}k_{\rm
    m}^\xi x^\xi K_{\alpha}(\xi,2\pi x k_{\rm
    m})\right)}{(\alpha-2)\Gamma\left(\frac{\alpha}{2}-1\right)},
\label{eq:sfan}
\ee 
where $\xi=\frac{\alpha}{2}-\frac{3}{2}$ and $K_{\alpha}$ is a
modified Bessel
function of the second kind.

\section[]{Relation between velocity dispersion along the
  line of sight and power spectrum}

$P_{\rm 1D}$ and $P_{\rm 3D}$ are related as $P_{\rm 1D}(k_z)=\int P_{\rm
  3D}(k_x,k_y,k_z)dk_x dk_y$, therefore eq. \ref{eq:sigPS} can be
re-written as
\be
\langle \sigma^2 (R)\rangle=\int_{-\infty}^{\infty}P_{\rm 1D}(k_{\rm z})(1-P_{\rm EM}(k_{\rm
  z})){\rm d}k_{\rm z},
\label{eq:sigma1d}
\ee
where $R$ is projected distance from the center.

If electron number density is described by $\beta$ model with
normalization $n_0$ and core radius $R_{\rm c}$
\be
n_{\rm e}(r)=\frac{n_0}{\left(1+\disp\frac{R^2+z^2}{R_{\rm c}^2}\right)^{\frac{3}{2}\beta}}
\ee
then the emissivity is
\be
\epsilon (r)=\frac{R_{\rm c}^{6\beta}}{(C+z^2)^{3\beta}},
\ee 
where $r^2=R^2+z^2$, $C=R_{\rm c}^2+R^2$ and we assume normalization $n_0=1$.
The Fourier transform of emissivity is
\be
\hat\epsilon(k_{\rm z})=\int_{-\infty}^{\infty}\frac{R_{\rm
    c}^{6\beta}}{(C+x^2)^{3\beta}}\cos(2\pi k_{\rm z} x){\rm d}x,
\label{fftem}
\ee
where terms with $i \sin(2\pi k_z z)$ are zero since integrand is symmetrical.
Dividing \ref{fftem} by the total flux and assuming $\beta >1/6$ one can
find weight $W(k_{\rm z})$ as
\be
W(k_{\rm z})=\frac{2C^\frac{\zeta}{2}k_{\rm z}^\zeta\pi^\zeta
  K_{\alpha}(-\zeta,2\sqrt{C}k_{\rm z}\pi)}{\Gamma(\zeta)},
\label{eq:weight}
\ee
where $\zeta=3\beta -\frac{1}{2}$ and $K_{\alpha}$ is a modified Bessel
function of the second kind.

\section[]{Ratio of observed RMS velocity to observed velocity
  dispersion}

Let us assume that spectrum is extracted from the region with the area
$\int\limits_{shape}dxdy$. The RMS velocity from this spectrum averaged over a number of
realizations is
\beq
&\nonumber \langle V^2_{\rm RMS}\rangle=\bigg\langle\frac{\int\limits_{shape}V^2_{\rm
    2D}(x,y)dxdy}{\int\limits_{shape}dxdy}\bigg\rangle-&\\&-\bigg\langle\left(\frac{\int\limits_{shape}V_{\rm
2D}(x,y)dxdy}{\int\limits_{shape}dxdy}\right)^2\bigg\rangle,&
\label{eq:vrms}
\eeq
 where $\langle\rangle$ denotes averaging over
 realizations. Accounting for eq. \ref{eq:v2d} we can re-write the
 first term in the above equation as
\beq
& \nonumber \bigg\langle \frac{\int\limits_{shape}\left(\int V_{\rm
    3D}(x,y,z)\epsilon(z)dz\right)^2dxdy}{\int\limits_{shape}dxdy}\bigg\rangle=&\\
&\nonumber= \disp\frac{1}{\int\limits_{shape}dxdy}\int\limits_{shape}\int\langle\hat V_{\rm
    3D}(k_{x_1},k_{y_1},k_{z_1})\rangle e^{i2\pi k_{x_1}x}e^{i2\pi
    k_{y_1}y}\times&\\&\nonumber \times F_{\rm EM}(k_{z_1}) \langle\hat V_{\rm
    3D}(k_{x_2},k_{y_2},k_{z_2})\rangle e^{i2\pi k_{x_2}x}e^{i2\pi
    k_{y_2}y}F_{\rm EM}(k_{z_2})\times&\\&\times d^3k_1d^3k_2 dxdy,&
\eeq
where $F_{\rm EM}$ is a Fourier transform of emissivity along the line
of sight.
Averaging over a number of realizations will leave non-zero terms only
if $k_1=k_2$. Therefore, the first term in eq. \ref{eq:vrms} is
\beq
&\nonumber \disp\frac{\int\limits_{shape}\int P_{\rm 3D}(k_x,k_y,k_z)P_{\rm
    EM}(k_z)d^3kdxdy}{\int\limits_{shape}dxdy}=&\\
&=\int P_{\rm 3D}(k_x,k_y,k_z)P_{\rm EM}(k_z)dk_xdk_ydk_z.&
\eeq

The second term in eq. \ref{eq:vrms} can be written as
\beq
&\bigg\langle\left(\frac{\int\limits_{shape}\int\limits_zV_{\rm
    3D}(x,y,z)\epsilon(z)dxdydz}{\int\limits_{shape}dxdy}\right)^2\bigg\rangle=&\\
&\nonumber =\bigg\langle\left(\frac{\int\limits_{shape}\int\hat V_{\rm
    3D}(k_x,k_y,k_z)e^{i2\pi xk_x}e^{i2\pi yk_y}F_{\rm EM}(k_z)d^3kdxdy}{\int\limits_{shape}dxdy}\right)^2\bigg\rangle.&
\eeq
Squaring and averaging over realizations yield
\beq
&\nonumber \frac{1}{\left(\int\limits_{shape}dxdy\right)^2}\int P_{\rm
  3D}(k_x,k_y,k_z)P_{\rm EM}(k_z)\times&\\
&\nonumber\times\int\limits_{shape} e^{-i2\pi k_{x_1}x_1}e^{-i2\pi
  k_{y_1}y_1}\times&\\
&\times\int\limits_{shape}e^{i2\pi k_{x_2}x_2}e^{i2\pi k_{y_2}y_2}dx_1dy_1dx_2dy_2d^3k,&
\eeq
or
\beq
&\nonumber \frac{1}{\left(\int\limits_{shape}dxdy\right)^2}\int P_{\rm
  3D}(k_x,k_y,k_z)P_{\rm EM}(k_z)\times&\\
&\nonumber\times\int S(x_1,y_1) e^{-i2\pi k_{x_1}x_1}e^{-i2\pi
  k_{y_1}y_1}\times&\\
&\times\int S(x_2,y_2)e^{i2\pi k_{x_2}x_2}e^{i2\pi k_{y_2}y_2}dx_1dy_1dx_2dy_2d^3k,&
\eeq
where $S(x,y)$ is a mask, which is zero outside and unity inside the
area, from which spectrum is extracted. Denoting power spectrum of
$S(x,y)/(\int\limits_{shape}dxdy)^2$ as $P_{\rm SH}(k_x,k_y)$, the
second term in eq. \ref{eq:vrms} becomes
\be
\int P_{\rm 3D}(k_x,k_y,k_z)P_{\rm EM}(k_z)P_{\rm SH}(k_x,k_y)dk_xdk_ydk_z.
\ee
Therefore, the final expression for the RMS of the projected velocity
field is
\be
\langle V^2_{\rm RMS}\rangle=\int P_{\rm 3D}(k_x,k_y,k_z)P_{\rm EM}(k_z)(1-P_{\rm SH}(k_x,k_y))d^3k.
\ee
And the ratio of $V^2_{\rm RMS}$ and velocity dispersion along
the line of sight is
\be
\frac{\langle V^2_{\rm RMS}\rangle}{\langle \sigma^2\rangle}=\frac{\int P_{\rm 3D}(k_x,k_y,k_z)P_{\rm EM}(k_z)(1-P_{\rm SH}(k_x,k_y))d^3k}{\int P_{\rm 3D}(k_x,k_y,k_z)(1-P_{\rm EM}(k_z))d^3k}.
\ee
\bsp
\label{lastpage}
\end{document}